\documentclass[amssymb,aps,twocolumn,floats,showpacs]{revtex4-1}
\usepackage{color,graphicx,float}
\usepackage{natbib}

\usepackage{fancybox}
\usepackage{graphicx}
\usepackage{tabularx}

\begin{document}

\title{Exotic Superconductivity in the Extended Attractive Hubbard Model}

\author{Swagatam Nayak}
\author{Sanjeev Kumar}

\address{
Indian Institute of Science Education and Research (IISER) Mohali, Sector 81, S.A.S. Nagar, Manauli PO 140306, India \\
}

\begin{abstract}

We show that the extended attractive Hubbard model on a square lattice hosts a variety of superconducting phases, including exotic mixed-symmetry phases with $d_{x^2-y^2} + {\rm i} [s + d_{x^2+y^2}]$ and $ d_{x^2-y^2} + p_{x}$ symmetries, and a novel $p_x + {\rm i} p_y$ state.
The calculations are performed within the mean-field Bogoliubov-deGennes (BdG) framework. The ground states of the BdG Hamiltonian are obtained via a minimization scheme that does not impose symmetry constraints on the superconducting solutions, hence allowing a mixing of $s$-, $p$- and $d$-wave order parameters.
Our results show that extended attractive Hubbard model can serve as an effective model for investigating properties of exotic superconducting states.

\end{abstract}
\date{\today}


\maketitle

\noindent
\underline{\it Introduction:} 
Identifying the symmetry of the superconducting (SC) order parameter (OP) is an important step towards understanding the properties of a SC state \cite{Tsuei2000}. The OP symmetry can also provide crucial insights regarding possible pairing mechanisms.
Indeed, the appearance of a non-$s$-wave component in the OP symmetry is taken as an indication of unconventional pairing mechanism. 
Nontrivial OP symmetries have been experimentally identified in many SC materials. 
Cuprates provide a famous example where the OP symmetry is known to be of $d$-wave type with a possible mixing of a secondary $s$-wave or $p$-wave component in some materials \cite{Mathai1995, Li1993, VanHarlingen1995, Muller1995, Betouras1999, Khasanov2007, Armitage2010}.
Recent ARPES experiments also show the evolution of the OP from a nodeless form to the nodal $d$-wave form \cite{Razzoli2012}.
An exotic chiral $p$-wave OP has been put forward as a strong candidate for the SC state in Sr$_2$RuO$_4$ \cite{Kallin2012,Kallin2016,Scaffidi2014,Luke1998}.
Spin triplet SC order has also been inferred from Knight-shift experiments on
Bechgaard salts, TMTSF$_2$PF$_6$ and TMTSF$_2$ClO$_4$ \cite{Belin1997,Lee1997,Lee2001,Lee2003,Oh2004}.
The possibility of mixed parity superconductivity, although not reported in any experiment yet, has not been ruled out \cite{Gorkov2001,Sergienko2004,Musaelian1996}.
Therefore, understanding and characterizing unconventional mixed-symmetry SC states remain a problem of critical importance.

It is well accepted that an effective attraction between electrons is a prerequisite for generating SC order. Therefore, effective models with attractive interactions
are commonly employed for investigating properties of SC states \cite{Micnas1988,Micnas1990,Keller2001a,Paiva2004}.
The simplest choice among such models is the attractive Hubbard model (AHM) which has been extensively studied using a variety of numerical and semi-analytical methods \cite{Anderson1975,Montorsi1996,Keller2001a,Capone2002,Singh1991,Gyorffy1991a,Freericks1993a,Allen2001a}. The on-site AHM allows for the conventional $s$-wave superconductivity. Including a nearest-neighbor (nn) attractive term readily supports a $d$-wave SC solution \cite{Arrachea1999,Mayr2006,Maiti2015,Micnas1988}.
It has been asserted that in some cases the induced attraction between electrons is not large enough to overcome the on-site Coulombic repulsion. However it can overcome the nn Coulombic repulsion, and therefore
an effective model with on-site repulsion and nn attraction may be realized \cite{Micnas1988,Hirsch1985b,Meintrup1995}. Indeed, this is a popular model for studying the competition between 
antiferromagnetism and d-wave superconductivity in the context of cuprates \cite{Micnas1988, Kuroki1994, Scalapino2012, Huang2013}.
Another realistic possibility is that the induced attraction overcomes both the on-site and nn repulsive interactions, leading to
an EAHM. Surprisingly, this model has not been explored much for the possibility of unconventional, particularly the mixed OP symmetry, SC solutions \cite{Mori1989, Huang2013}. 

In this work, we unveil the exciting possibility of the existence of unconventional mixed symmetry SC states in an EAHM on a square lattice. A justifiable approximation on the nn attractive interaction followed by a general decoupling scheme together with an explicit minimization procedure allows us to construct comprehensive phase diagrams for the model.
Superconducting phases with mixed OPs dominate the phase diagram. 
We present simple energetic arguments for the stability of mixed OP phases. Two of the unconventional phases, the chiral $p_x + {\rm i}p_y$ order and the $d_{x^2-y^2} + p_x$ order, exist over a wide parameter regime. Both these orderings also support non-trivial edge-state dispersions. While the two OP symmetries mentioned above are directly relevant to some cuprates and Sr$_2$RuO$_4$, respectively, our results have a general implication that the EAHM can be a universal effective model for studying unconventional superconductivity, just as the on-site attractive Hubbard model serves this purpose for conventional $s$-wave superconductivity.

\noindent
\underline{\it Extended Attractive Hubbard Hamiltonian:} 
We begin with the EAHM defined on a 2D square lattice. The model is described by the Hamiltonian,

\begin{eqnarray}
H & = &  - t \sum_{\langle ij\rangle,\sigma} [ c^\dagger_{i\sigma} c^{}_{j\sigma} + H.c.] -
\mu \sum_{i\sigma}c^\dagger_{i\sigma} c^{}_{i\sigma} \nonumber \\
& & -U \sum_{i} n_{i\uparrow}n_{i\downarrow} - V \sum_{\langle ij \rangle} n_{i}n_{j}.
\label{Ham}
\end{eqnarray}
\noindent
Here $c_{i\sigma} (c_{i\sigma}^\dagger$) annihilates (creates) an electron
at site ${i}$ with spin $\sigma$, $\langle ij \rangle$ implies that sites
$i$ and $j$ are nearest neighbors. $\mu$ is the chemical
potential, $n_{i\sigma}=c_{i\sigma}^{\dagger}c^{}_{i\sigma}$ is the electron number operator at site
$i$ and spin-projection $\sigma$, and $n_i = n_{i \uparrow} + n_{i \downarrow}$. $U$ and $V$ denote the strengths of
on-site and nearest neighbor attractive interactions, respectively.
Using $t=1$ as the basic energy scale, and restricting ourselves to zero temperatures ($T=0$), we are left with three independent parameters in the Hamiltonian, {\it viz.}, $U$, $V$ and $\mu$. 

We analyze the Hamiltonian in Eq.~(\ref{Ham}) by making a mean-field approximation, also known as the Bogoliubov-deGennes (BdG) approximation, for 
the interaction term \cite{deGennes1999}.
In the intersite attractive term we ignore the same-spin attraction parts $n_{i \uparrow} n_{j \uparrow}$ and $n_{i \downarrow} n_{j \downarrow}$. 
This can be qualitatively justified for systems where superconductivity emerges in the vicinity of antiferromagnetism. Apparently, the antiferromagnetic tendency
ensures that oppositely spin-oriented electrons are more likely to reside on neighboring sites as compared to those with same spin orientation.
The mean-field Hamiltonian is obtained by making replacements $c^{\dagger}_{i\uparrow} c^{\dagger}_{j\downarrow} \rightarrow \langle c^{\dagger}_{i\uparrow} c^{\dagger}_{j\downarrow} \rangle + \hat{\delta}^{\dagger}$ and $c^{}_{j\downarrow} c^{}_{i\uparrow} \rightarrow \langle c^{}_{j\downarrow} c^{}_{i\uparrow} \rangle + \hat{\delta}^{}$ and ignoring terms that are bilinear in $\hat{\delta}$.
This leads to the BdG Hamiltonian,

\begin{eqnarray}
H_{\rm{BdG}} & = & - t \sum_{\langle ij \rangle,\sigma} \left[ c_{i\sigma}^\dagger
c_{j\sigma} + H.c. \right] - U \sum_{i}\left[\Delta_{i}c_{i\uparrow}^\dagger c_{i\downarrow}^\dagger + H.c. \right] \nonumber \\
& & - V \sum_{i \gamma} \left[\Delta^{+}_{i, \gamma}c_{i\uparrow}^\dagger c_{i+\gamma \downarrow}^\dagger + \Delta^{-}_{i \gamma}c_{i-\gamma \downarrow}^\dagger c_{i \uparrow}^\dagger
+ H.c. \right] \nonumber \\
& & +U \sum_i | \Delta_i |^2 + V \sum_{i \gamma} \left[ |\Delta^+_{i\gamma}|^2 + |\Delta^-_{i\gamma}|^2  \right]
.
\label{BdGequation}
\end{eqnarray}        
\noindent
In the above we have introduced the pair expectation values in the ground state as, $\Delta_i = \langle c_{i\downarrow} c_{i\uparrow} \rangle$, $\Delta^+_{i,\gamma} = \langle c_{i+\gamma \downarrow} c_{i\uparrow} \rangle$, and 
$\Delta^-_{i,\gamma} = \langle c_{i-\gamma \downarrow} c_{i\uparrow} \rangle$, where $\gamma$ denotes the unit vectors $+\hat {\bf x}$ and $+\hat {\bf y}$ on the square lattice.
Note that we do not impose the commonly used spin-singlet symmetry constraint on the pair expectation values, and therefore, in general, $\Delta^+_{i, \gamma} \neq \Delta^-_{i+\gamma, \gamma}$ (see supplemental material). 
For simplicity, we focus on the SC phases that respect the translational symmetry of the Hamiltonian. Hence, we assume the above quantum expectation values to be independent of lattice sites. Going over to the Fourier space by using, $c_{i \sigma} = N_s^{-1/2} \sum_{\bf k} e^{-{\rm i} {\bf k} \cdot {\bf r}_i} c_{{\bf k} \sigma}$ and $c^{\dagger}_{i \sigma} = N_s^{-1/2} \sum_{\bf k} e^{{\rm i} {\bf k} \cdot {\bf r}_i} c^{\dagger}_{{\bf k} \sigma}$, $N_s$ being the number of sites, the Hamiltonian can be reduced to a $2 \times 2$ matrix form. The resulting mean-field Hamiltonian in the Nambu spinor notation is,

\begin{eqnarray}
H_{MF} & = & \sum_{\bf k} \left[ \begin{array}{c c} c^{\dagger}_{{\bf k} \uparrow} & c^{}_{-{\bf k} \downarrow} \end{array} \right]
    \left[ \begin{array}{c c} h_{11}({\bf k})  &  h_{12}({\bf k}) \\ h_{21}({\bf k}) & h_{22}({\bf k}) \end{array} \right] \left[ \begin{array}{c} c^{}_{{\bf k} \uparrow} \\                                                                                               
    c^{\dagger}_{-{\bf k} \downarrow} \end{array}  \right] + \nonumber \\ 
  & & N \left \{ U |\Delta|^2 + V (|\Delta^+_x|^2 + |\Delta^-_x|^2 + |\Delta^+_y|^2 + |\Delta^-_y|^2) \right \}. \nonumber \\
\label{Ham-mf}  
\end{eqnarray}

\noindent
The matrix elements in the above equation are explicitly given by,
\begin{eqnarray}
h_{11}({\bf k}) & = & -2t(\cos k_x+\cos k_y) - \mu = -h_{22}({\bf k}) \nonumber  \\
h_{12}({\bf k}) & = & -U\Delta - V(\Delta^+_x e^{-{\rm i}k_x} + \Delta^-_x e^{{\rm i}k_x} \nonumber \\ 
          &   & + \Delta^+_y e^{-{\rm i}k_y} + \Delta^-_y e^{{\rm i}k_y}) = h^*_{21}({\bf k}).
\end{eqnarray}

The electronic part of $H_{mf}$ can be diagonalized for an arbitrary set of mean-field parameters using Bogoliubov transformations. Therefore, the problem now reduces to finding the set $\{\Delta \} \equiv \{ \Delta_0, \Delta^+_x, \Delta^-_x, \Delta^+_y, \Delta^-_y \}$ that minimizes the total energy. We want to emphasize here that in most previous studies a particular form of the SC OP is assumed {\it a priori} \cite{Mori1989}. In contrast, we allow for all possible combinations of OPs and rely on energetics to pick the most stable SC order.

\noindent 
\underline{\it Minimization scheme:}
To put our results in proper context, we observe the following relations between the pair expectation values defined above the and the commonly used SC OPs.

\begin{eqnarray}
 \Delta_s & = & \Delta_0  \nonumber \\
 \Delta_{d_{x^2+y^2}} & = & (\Delta^+_x + \Delta^-_x + \Delta^+_y + \Delta^-_y )/4 \nonumber \\
 \Delta_{d_{x^2-y^2}} & = & (\Delta^+_x + \Delta^-_x - \Delta^+_y - \Delta^-_y )/4 \nonumber \\
 \Delta_{p_x} & = & (\Delta^+_x - \Delta^-_x)/2 \nonumber \\
 \Delta_{p_y} & = & (\Delta^+_y - \Delta^-_y)/2.
\label{OP-def}
\end{eqnarray}

\noindent
The $s$-, $p$- and $d$-wave OPs defined above have their usual meaning. It is easy to see that the form-factors that enter the ${\bf k}$-space matrix acquire their typical pure-singlet or pure-triplet form in the limiting cases (see supplemental material). 
In addition to determining the magnitude of the OPs in the minimum energy state, we also need to take into account the relative phase angles between different OPs in the mixed states. Therefore, we carry out variational calculations for energy as a function of relative phase angle between different OPs. This helps us in reducing the number of variational parameters by fixing some of the relative phase angles (see supplemental material). 

\begin{figure}[t!]
\includegraphics[width=.98 \columnwidth,angle=0,clip=true]{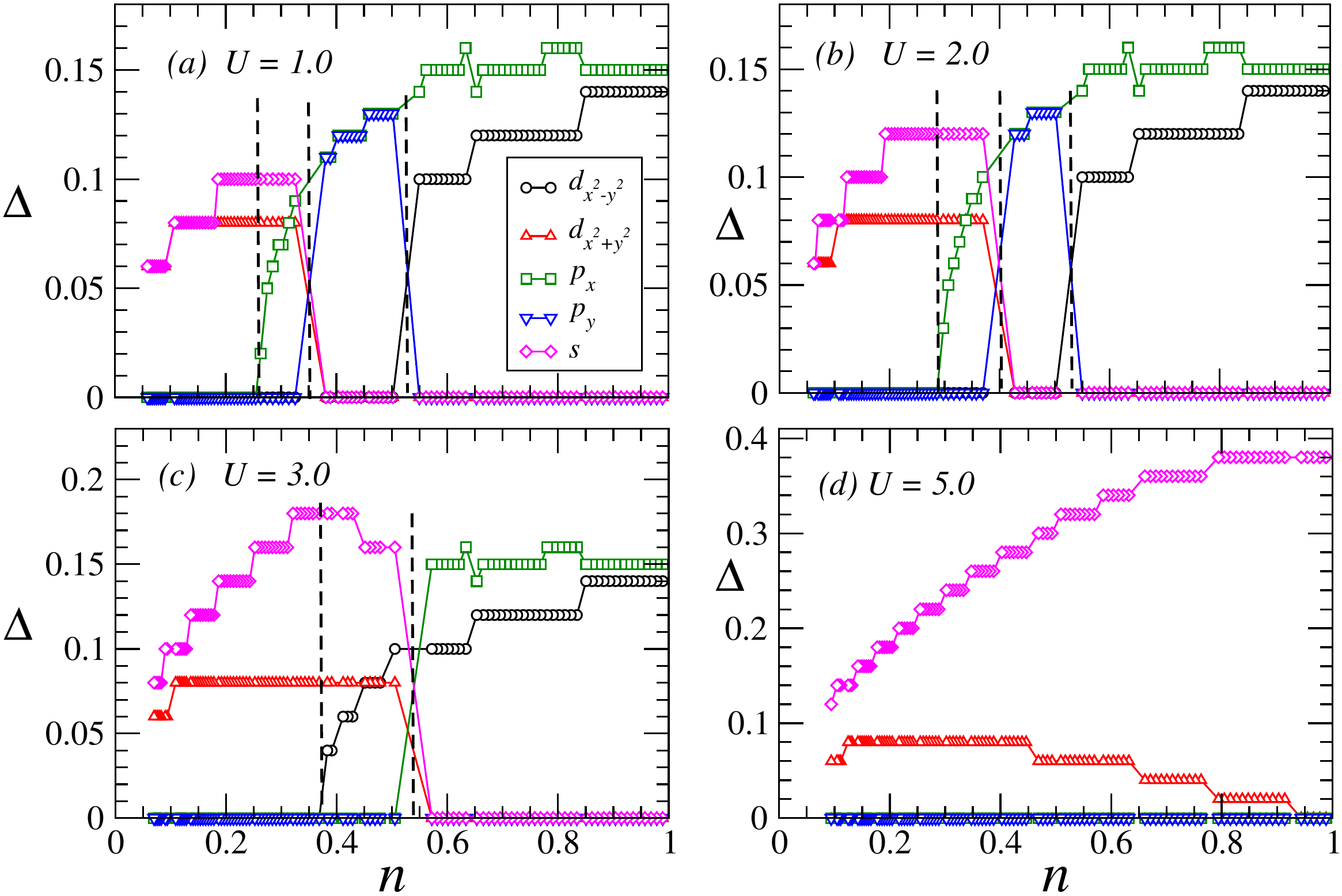}
\caption{(Color online) ($a$)-($d$) The values of various OPs corresponding to the minimum energy states as a function of average electronic density $n$. Results for representative values of $U$ are shown. The dashed vertical lines in each plot mark the boundary between qualitatively distinct phases. The results are obtained for $V = 4t$.
}
\label{fig1}
\end{figure}

\noindent
\underline{\it Order parameters and phase diagram:} 
We focus our discussion on the variations in $U/t$ and $\mu$ for a fixed value of $V/t = 4$. 
Direct minimization is carried out by varying different real-valued OPs and relative phase factors among them. The density dependence of SC OPs corresponding to minimum total energy are plotted in Fig. \ref{fig1}($a$)-($d$). For small $U$, the high-density regime is dominated by $d_{x^2-y^2}$ and $p_x$ OPs. Both $p_x$ and $p_y$ are finite in the intermediate density range. At further lower densities OPs with $p_x$, $s$ and $d_{x^2+y^2}$ symmetries are finite. Eventually, the low-density regime supports $s$ and $d_{x^2+y^2}$ OPs (see Fig. \ref{fig1}($a$)-($b$)). 
For larger values of on-site attraction, $\Delta_{p_x}$ and $\Delta_{p_y}$ remain zero, and instead $\Delta_{x^2+y^2}$ and $\Delta_{x^2-y^2}$ together with $s$-wave OP become finite (see Fig. \ref{fig1}($c$)). Finally, in the limit of large $U$, $s$-wave OP dominates, and extended-$s$ or $d_{x^2+y^2}$ OP is always finite.
We simultaneously track the values of relative phase angles between these OPs in the minimum energy state, allowing us to describe the specific combination of the mixed SC OPs. 

\begin{figure}[t!]
\includegraphics[width=.90\columnwidth,angle=0,clip=true]{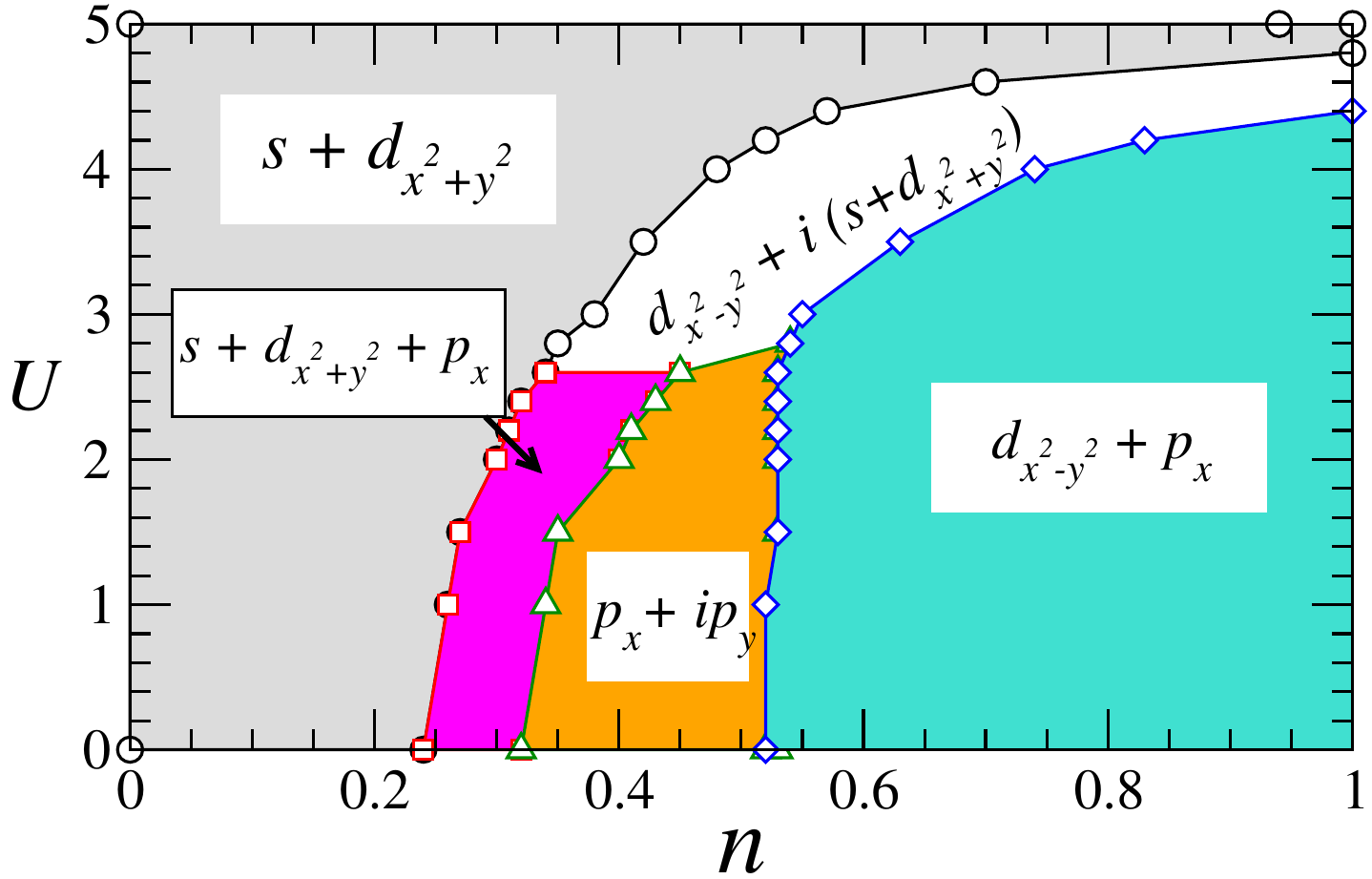}
\caption{(Color online) Phase diagram in the $U$-$n$ parameter space for intersite attraction strength $V = 4 t$. The results are obtained via brute-force minimization 
of total energy using a $16 \times 16$ $~{\bf k}$-point grid for different combinations of OPs described in text.}
\label{fig2}
\end{figure}

We summarize the results in the form of a $n-U$ phase diagram in Fig. \ref{fig2}. 
Most notably, a chiral $p_x + {\rm i}p_y$ order is present in the density range $0.35 < n < 0.55$ in the limit of weaker on-site attraction. Within this interaction regime, $p_x$ order also mixes with $d_{x^2-y^2}$ and extended $s$-wave order for different electronic densities. The possibility of chiral $p$-wave order in the extended repulsive Hubbard model has been pointed out within fluctuation exchange approximation \cite{Onari2004,Onari2005}.
The $d_{x^2-y^2} + p_x$ order is particularly stable over a large density regime.
This is intriguing as a number of experiments on cuprates report on the possibility of a secondary unconventional OP in addition to the dominant $d_{x^2-y^2}$ order. The secondary OP is proposed to be either $s$-wave or $p$-wave. Interestingly, phases with $d_{x^2-y^2}+{\rm i} [s+d_{x^2+y^2}]$ and $d_{x^2-y^2} + p_x$ OPs reside next to each other in the doping regime $0.6 < n < 1$ depending on the value of $U$. Moreover, for smaller values of $V$, we also find a pure $d_{x^2-y^2}$ order in the doping range $0.7 < n < 0.9$ (not shown here).
Note that even in the $U=0$ limit a finite $s$-wave order is introduced via $d_{x^2+y^2}$ ordering originating from the nn attractive term. For a range of $U$ values, the variations of $n$ with $\mu$ display step-like jumps with $\delta n \sim 0.03-0.05$. Therefore, the presence of a few phase separation pockets in the $n-U$ phase diagram cannot be ruled out. The mean-field Hamiltonian (\ref{Ham-mf}) can also be solved via the standard self-consistent approach \cite{deGennes1999}. We have checked that various SC states discussed above are also the self-consistent states (see supplemental material).
It is important to mention that we have not considered the competition of magnetic and charge-density-wave ordering in this study since our focus was on studying the competition between SC orders with different symmetries. Some of the SC phases, especially those near commensurate fillings, may be destabilized by the competing magnetic and charge ordering tendencies of the EAHM. However, it is also known that superconductivity is relatively more robust against quenched disorder compared to, for example, charge ordering \cite{Huscroft1997a}. Therefore, in a realistic situation where quenched disorder is always present, the above phase diagram will be of relevance.

\noindent
\underline{\it Bulk and edge-state spectra:}
We now investigate further some of the mixed OP SC phases discussed above. 
We begin with the calculation of tunneling density of states (TDOS) in different phases. Normalized TDOS is defined as,

\begin{equation}
 N(\omega) = 1/N_s \sum_{{\bf k}} |u_{{\bf k}}|^2 \delta(\omega - E_{\bf k}) + |v_{{\bf k}}|^2 \delta(\omega + E_{\bf k}),
\end{equation}

\noindent
where $E_{\bf k}$ is the energy dispersion for Bogoliubov quasiparticles and $u_{{\bf k}}$ measures electron-like amplitude in the quasiparticle state labeled by wave vector ${\bf k}$.
TDOS can be directly probed by tunneling experiments and therefore characterization of different mixed OP states in terms of TDOS is desirable \cite{Fischer2007, Balatsky2006}.

\begin{figure}[t!]
\includegraphics[width=.98\columnwidth,angle=0,clip=true]{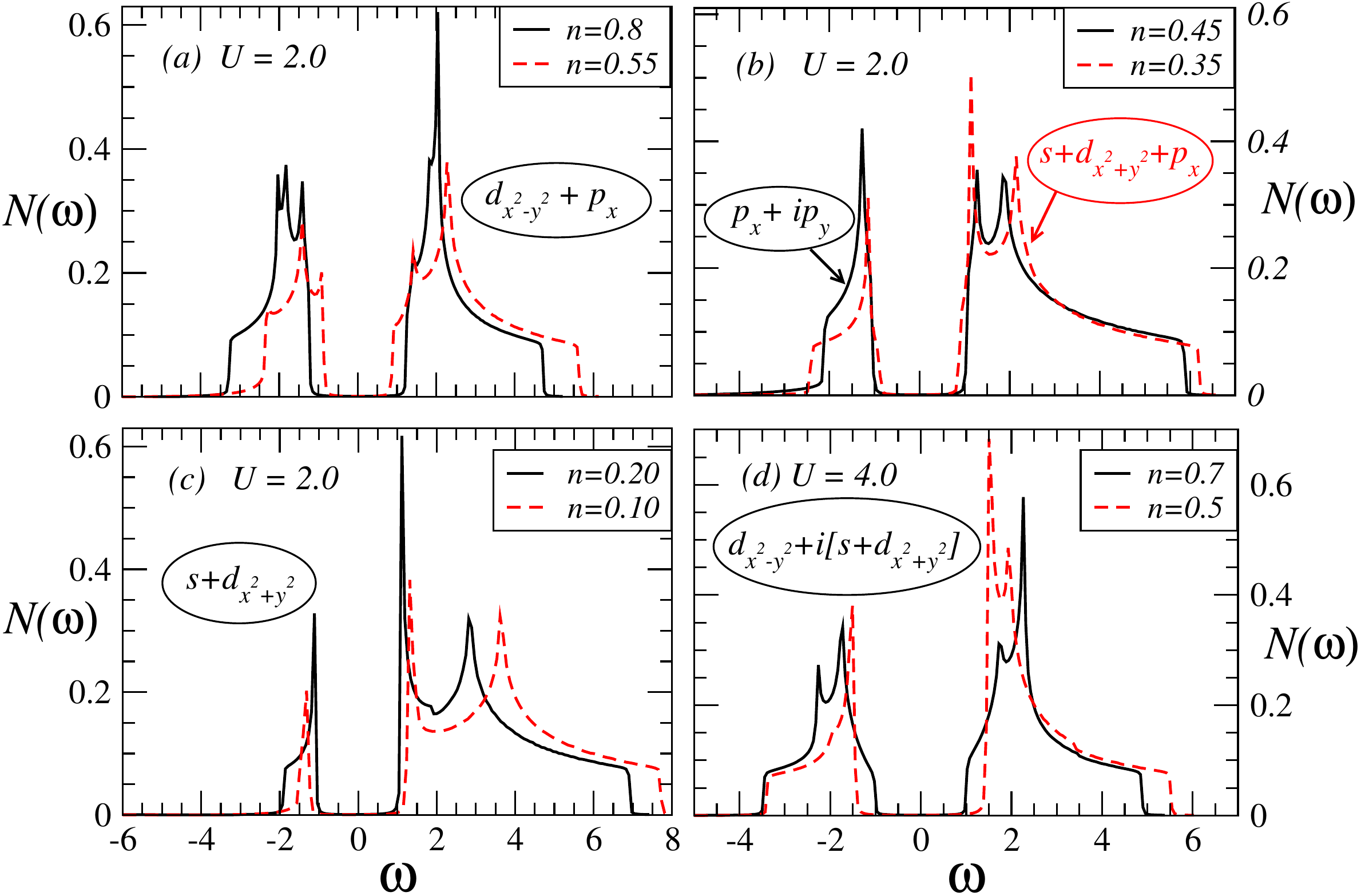}
\caption{(Color online) ($a$)-($d$) Density of states for electrons in different phases at different values of average electronic density. These calculations are performed on $600 \times 600$ ${\bf k}$-point grid. A Lorentzian broadening of $0.01 t$ is used.}
\label{fig3}
\end{figure}

Mixing of a $p_x$ component in the $d_{x^2-y^2}$ superconductivity completely modifies the TDOS structure and opens a clean gap much like that present in the simple $s$-wave superconductors (see Fig. \ref{fig3} ($a$)). Indeed, the nodes present in the $d_{x^2-y^2}$ gap function are removed by the presence of $ {\rm i} \Delta_{p_x}\sin k_x$ term. Multiple coherence peaks in the TDOS are also clearly observed. In fact, it is easy to see why a mixing of $p$-wave component is energetically favored. The system gains energy by pushing the eigenenergies further away from the chemical potential by opening a clean gap. The chiral $p$-wave order and the mixed $s+d_{x^2+y^2}+p_x$ orders also support a clean gap in the TDOS (see Fig. \ref{fig3} ($b$)). The $s + d_{x^2+y^2}$ ordering shows the expected TDOS with the coherence peaks residing right at the gap edge. In the $d_{x^2-y^2} + {\rm i}[s+d_{x^2+y^2}]$ state the features corresponding to $s$-wave and $d$-wave ordering are present at larger value of electronic density (see Fig. \ref{fig3} ($d$)). For the smaller density, the $d$-wave component reduces and the TDOS appears $s$-wave-like. The occurrence of a $d+is$ phase in extended Hubbard model has also been reported previously \cite{Maiti2015}. The present model can be used to fit tunneling data of unconventional superconductors in order to identify possible mixed OP symmetries.

\begin{figure}[t!]
\includegraphics[width=.98\columnwidth,angle=0,clip=true]{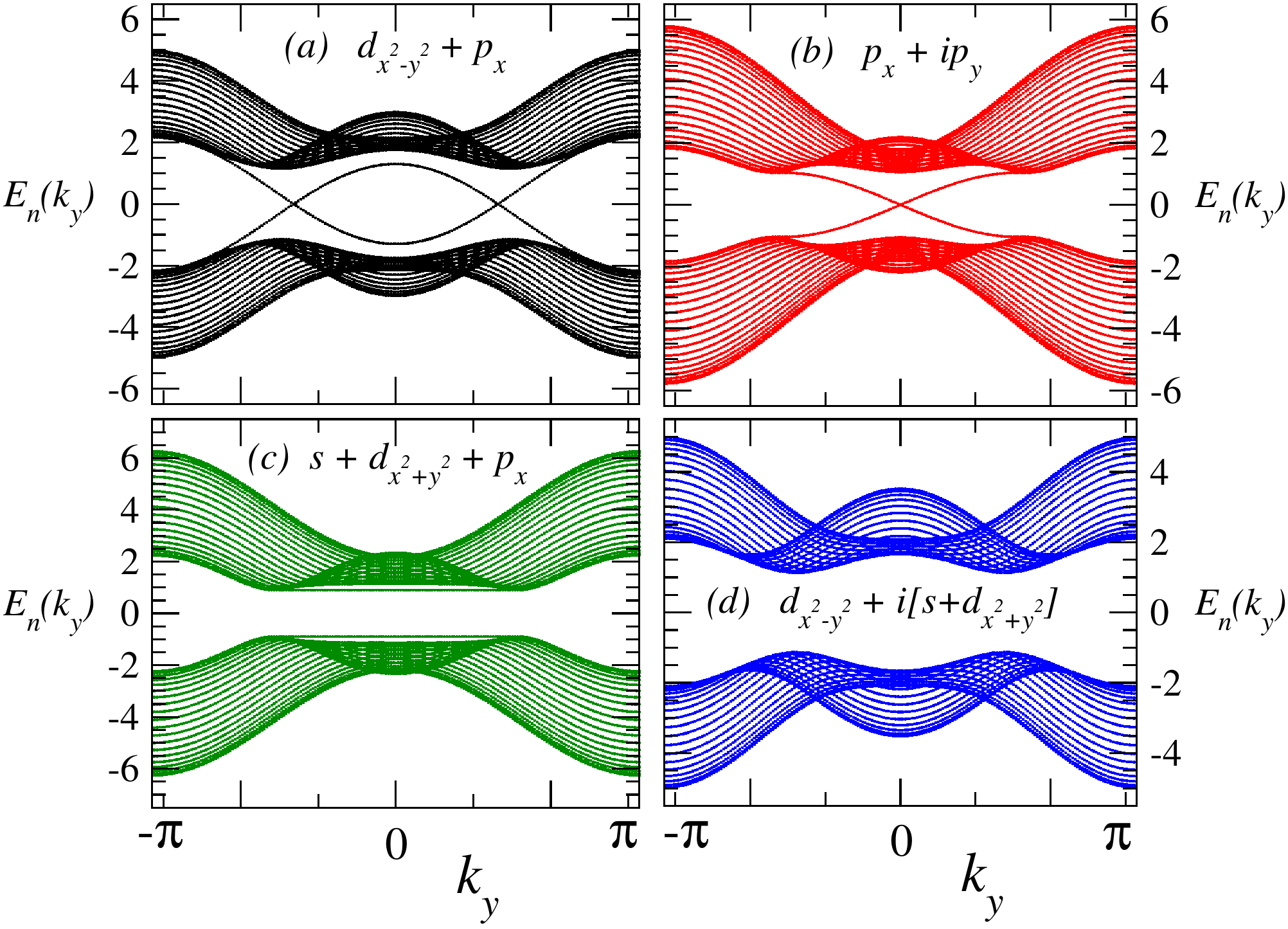}
\caption{(Color online) ($a$)-($d$) Bogoliubov quasiparticle dispersions for different SC states obtained 
by using open (periodic) boundary condition along $x$ ($y$) direction. Edge states disperse across the SC gap for, ($a$) $p_x + i p_y$  and for ($b$) $d_{x^2-y^2} + p_x$  OP symmetries. ($c$) and ($d$) display trivial gaps with no mid-gap edge states.}
\label{fig4}
\end{figure}

We further investigate the nature of various phases in terms of their edge-state spectra. To this end, we perform calculations on a $20\times200$ stripe by imposing periodic boundary conditions only along the $y$ direction and plotting the spectra as a function of $k_y$. Two of the new phases obtained from our calculations turn out to be trivial as no edge states are found to appear inside the SC gap (see Fig. \ref{fig4}($c$)-($d$)). The chiral $p$-wave superconductor shows the expected non-trivial behavior wherein counter-propagating edge states appear in the gap (see Fig. \ref{fig4}($b$)) \cite{Taylor2012}. An intriguing situation occurs for $d_{x^2-y^2}+p_x$ superconductor where pairs of states are present on each edge (see Fig. \ref{fig4}($a$)). 
While the topology of the bulk band will remain trivial in this case as the states traverse back to their respective original band, the presence of such mid-gap states will have observable consequences. Indeed, if such a situation can be realized in a real material, then the surface spectroscopy with voltage bias smaller than the gap value will have contributions from the edge states. This is in contrast to the situation where such states are absent, and only Andreev reflection contributions are observed in tunneling.

\noindent
\underline{\it Conclusion:} 

In conclusion, we have shown in this work that the EAHM treated without imposing symmetry constraints presents an exciting possibility for hosting a variety of SC states with mixed OP symmetries. Our approach allowed for competition between SC orders of $s$- $p$- and $d$-wave type.
The resulting phase diagram hosts some very interesting and new SC phases. Most notable of these are, (i) the chiral $p$-wave state, (ii) states with mixed $d$-wave and $p$-wave, and with $s$-wave, extended $s$-wave and $p_x$ symmetries, and (iii) a $d_{x^2-y^2}+ {\rm i}[s+d_{x^2+y2}]$ SC phase. To the best of our knowledge, the possibility of such mixed order-parameter phases has not been explored in the EAHM \cite{Monthoux1999}. Some experiments on cuprates report the possibility of a mixed $s$ and $d$ wave order \cite{Muller1995}, while a possible mixing of a $p$-wave component with the $d$-wave order has been inferred via thermal transport measurements \cite{Movshovich1998}. Experiments on Sr$_2$RuO$_4$ indicate a chiral $p_x + {\rm i}p_y$-wave order. Similarly, $p$-wave SC OP is consistent with experiments on Bechgaard salts TMTSF$_2$PF$_6$ and TMTSF$_2$ClO$_4$ \cite{Brown2015,Belin1997,Lee1997,Lee2001,Lee2003,Oh2004}. Although a microscopic theory of superconductivity in some of these systems is still awaited, our results suggest that EAHM can serve as the effective model for a variety of superconductors. This is in the same spirit as the on-site attractive Hubbard model is the phenomenological model for $s$-wave superconductivity. Indeed, the effect of non-magnetic and magnetic impurities, influence of Zeeman and Peierls's terms arising from an external magnetic field, effects of next-nearest hopping, etc. are some of the problems that can be readily addressed using the present model. The model can be made material specific by estimating the values of effective on-site and nn electron-electron attractions. Such model studies can help in 
a microscopic characterization of various mixed-symmetry states and can be useful in improving our understanding of the rich experimental data available on unconventional superconductors.

\noindent
\underline{\it Acknowledgments:}
We acknowledge the use of High-Performance Computing Facility at IISER Mohali.

\section{APPENDICES}
\subsection{General Hartree-Fock Decoupling in the Pairing Channel }

Here we discuss in detail how a general decoupling of the nearest-neighbor (nn) attractive interaction allows for possible mixed order parameter solutions.
The interaction term is given by,
\begin{equation}
H_{\rm{int}} = -U \sum_{i} n_{i\uparrow}n_{i\downarrow} - V \sum_{\langle ij \rangle} n_{i}n_{j}.
\label{Ham-int}
\end{equation}

\noindent
The Hartree-Fock decoupling in the pairing channel of the first term in Eq. (\ref{Ham-int}) is straightforward, and leads to the replacement $n_{i\uparrow}n_{i\downarrow}  \longrightarrow 
[ \langle c^{}_{i \downarrow} c^{}_{i \uparrow} \rangle c^{\dagger}_{i \uparrow} c^{\dagger}_{i \downarrow} + \langle c^{\dagger}_{i \uparrow} c^{\dagger}_{i \downarrow} \rangle
c^{}_{i \downarrow} c^{}_{i \uparrow} - \langle c^{}_{i \downarrow} c^{}_{i \uparrow} \rangle \langle c^{\dagger}_{i \uparrow} c^{\dagger}_{i \downarrow} \rangle].
$
The second term can be written as,
\begin{equation}
H^{nn}_{\rm{int}} = -V \sum_{i,\gamma = +\hat{\bf{x}},  +\hat{\bf{y}}}  (n_{i\uparrow} + n_{i\downarrow})(n_{i+\gamma,\uparrow} + n_{i+\gamma,\downarrow})
\end{equation}
Expanding further, we obtain four terms corresponding to each $i, i+\gamma$ bond. These are $n_{i\uparrow} n_{i+\gamma,\uparrow}$, $n_{i\downarrow} n_{i+\gamma,\downarrow}$, $n_{i\uparrow} n_{i+\gamma,\downarrow}$ and 
$n_{i\downarrow} n_{i+\gamma,\uparrow}$. We assume that electrons with identical spin orientations are less likely to reside on nn sites, and taking an approximation we drop the $\uparrow \uparrow$ and $\downarrow \downarrow$ interaction terms altogether. Rearranging the order of $c$ operators, we can write these as,
\begin{equation}
H^{nn}_{\rm{int}} \approx -V \sum_{i,\gamma = +\hat{\bf{x}},  +\hat{\bf{y}}}  [ c_{i\uparrow}^\dagger c_{i+\gamma \downarrow}^\dagger c_{i+\gamma \downarrow}^{} c_{i\uparrow}^{} + 
 c_{i+\gamma \uparrow}^\dagger c_{i\downarrow}^\dagger c_{i\downarrow}^{} c_{i+\gamma \uparrow}^{} ]
\end{equation}

Implementing the Hartree-Fock decoupling in the pairing channel, we find,

\begin{eqnarray}
H^{nn}_{\rm{int}} & \approx & -V \sum_{i,\gamma}  [ c_{i\uparrow}^\dagger c_{i+\gamma \downarrow}^\dagger \langle c_{i+\gamma \downarrow}^{} c_{i\uparrow}^{} \rangle 
 + \langle c_{i\uparrow}^\dagger c_{i+\gamma \downarrow}^\dagger \rangle c_{i+\gamma \downarrow}^{} c_{i\uparrow}^{} \nonumber \\ 
 & & - \langle c_{i\uparrow}^\dagger c_{i+\gamma \downarrow}^\dagger\rangle  \langle c_{i+\gamma \downarrow}^{} c_{i\uparrow}^{} \rangle ] +
[c_{i+\gamma \uparrow}^\dagger c_{i\downarrow}^\dagger \langle c_{i\downarrow}^{} c_{i+\gamma \uparrow}^{} \rangle \nonumber \\ 
& & + \langle c_{i+\gamma \uparrow}^\dagger c_{i\downarrow}^\dagger \rangle c_{i\downarrow}^{} c_{i+\gamma \uparrow}^{} - 
\langle c_{i+\gamma \uparrow}^\dagger c_{i\downarrow}^\dagger \rangle \langle c_{i\downarrow}^{} c_{i+\gamma \uparrow}^{} \rangle ].
\end{eqnarray}

\noindent
Note that in order to retain the generality of the decoupling we have introduced two different pair expectation values for a given nn pair of sites. These expectation values, $\Delta^{+}_{i,\gamma} = \langle c_{i+\gamma \downarrow}^{} c_{i\uparrow}^{} \rangle$ and 
$ \Delta^{-}_{i+\gamma,\gamma}=\langle c_{i\downarrow}^{} c_{i+\gamma \uparrow}^{} \rangle$ need not be equal, in principle. Indeed, if we assume that the pair satisfies antisymmetry under spin exchange, then 
$\Delta^{+}_{i,\gamma} = \Delta^{-}_{i+\gamma,\gamma}$, and if the pair satisfies antisymmetry under site-index exchange then $\Delta^{+}_{i,\gamma} = -\Delta^{-}_{i+\gamma,\gamma}$. In most studies a singlet condition on the pairing correlations is imposed and therefore the possibility of odd parity pairing in this model is left out. Here, we do not impose this symmetry constraint on our pairing correlations.

\noindent
The electronic part of the mean-field Hamiltonian reads,

\begin{eqnarray}
H_{el} & = & \sum_{\bf k} \left[ \begin{array}{c c} c^{\dagger}_{{\bf k} \uparrow} & c^{}_{-{\bf k} \downarrow} \end{array} \right]
    \left[ \begin{array}{c c} h_{11}({\bf k})  &  h_{12}({\bf k}) \\ h_{21}({\bf k}) & h_{22}({\bf k}) \end{array} \right] \left[ \begin{array}{c} c^{}_{{\bf k} \uparrow} \\                                                                                               
    c^{\dagger}_{-{\bf k} \downarrow} \end{array}  \right],
\label{Ham-mf}  
\end{eqnarray}

\noindent
where the matrix elements are specified as,
\begin{eqnarray}
h_{11}({\bf k}) & = & -2t(\cos k_x+\cos k_y) - \mu = -h_{22}({\bf k}) \nonumber  \\
h_{12}({\bf k}) & = & -U\Delta_0 - V(\Delta^+_x e^{-{\rm i}k_x} + \Delta^-_x e^{{\rm i}k_x} \nonumber \\ 
          &   & + \Delta^+_y e^{-{\rm i}k_y} + \Delta^-_y e^{{\rm i}k_y}) = h^*_{21}({\bf k}).
\end{eqnarray}

\noindent
Note that we will recover the standard form of $h_{11}({\bf k})$ if we assume different symmetry relations between $\Delta^{+}_{\gamma}$ and $\Delta^{-}_{\gamma}$. For spin singlet pairing we get
$\Delta^{+}_{\gamma} = \Delta^{-}_{\gamma}$, and $\cos k_x + e^{{\rm i} \Phi} \cos k_y$ form is obtained. Similarly, imposing spin triplet symmetry $\Delta^{+}_{\gamma} = - \Delta^{-}_{\gamma}$, and we find the ${\rm i}(\sin k_x + e^{{\rm i} \Phi} \sin k_y)$ form. 
For a given set $\{\Delta \} \equiv \{ \Delta_0, \Delta^+_x, \Delta^-_x, \Delta^+_y, \Delta^-_y \}$, we can diagonalize the Hamiltonian Eq. (\ref{Ham-mf}) via the Bogoliubov transformations,

\begin{equation}
\left[ \begin{array}{c} c^{}_{{\bf k} \uparrow}  \\                                                                                         
    c^{\dagger}_{-{\bf k} \downarrow} \end{array}  \right] = \left[ \begin{array}{c c}  u_{\bf k} &  -v^*_{\bf k}  \\  v_{\bf k} &  u^*_{\bf k}  \end{array}  \right]
\left[ \begin{array}{c} \gamma^{}_{{\bf k} 0} \\                                                                                              
    \gamma^{\dagger}_{-{\bf k} 1} \end{array} \right],
\label{BogT}
\end{equation}

\noindent
where $u_{\bf k}$ and $v_{\bf k}$ are complex numbers satisfying $ |u_{\bf k}|^2 + |v_{\bf k}|^2 = 1$ for all ${\bf k}$, and $\gamma$, $\gamma^{\dagger}$ are the annihilation and creation operators for Bogoliubov quasiparticles. The resulting quasiparticle dispersion is given by,

\begin{eqnarray}
 E_{{\bf k}} & = & \sqrt{ (-2t(\cos k_x+\cos k_y) - \mu)^2 + \Delta^2_{g}}, \nonumber \\
 \Delta^2_g & = & \vert -U\Delta_0 - V(\Delta^+_x e^{-{\rm i}k_x} + \Delta^-_x e^{{\rm i}k_x} \nonumber \\ 
          &   & +\Delta^+_y e^{-{\rm i}k_y} + \Delta^-_y e^{{\rm i}k_y}) \vert^2 
\end{eqnarray}

\noindent
Using the above quasiparticle spectrum along with the purely classical terms in the mean-field Hamiltonian Eq. (3) of the main text, we can compute the total energy of any general state
specified by a set $\{ \Delta \}$. Therefore, it is now a simple exercise to minimize the total energy w.r.t. the set $\{ \Delta \}$ of pairing correlations.


\subsection{Relative phase angles between different order parameters}

\begin{figure}[H]
\includegraphics[width=\columnwidth]{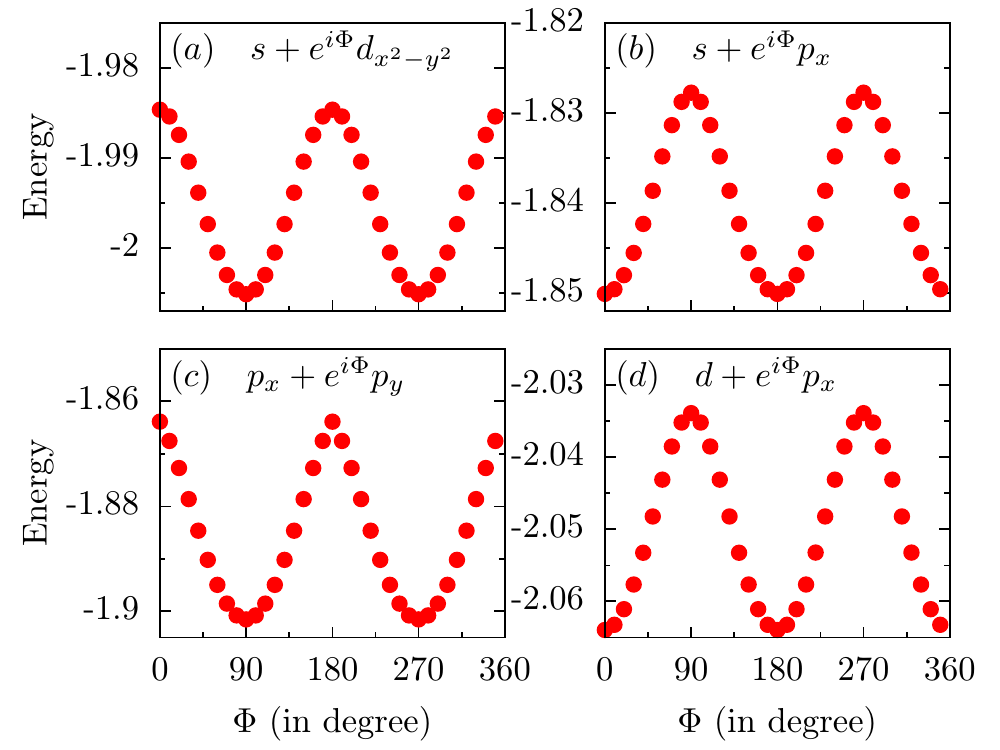}
\caption{Variation of energy with relative phase angle ($\Phi$) between two finite order parameters, at $\mu=0$ ($n=1.0$), chosen pair-wise for the following cases : (a) $\Delta_s+e^{{\rm i} \Phi}\Delta_{d_{x^2 -y^2}}$, (b) $\Delta_s+e^{{\rm i} \Phi}\Delta_{p_x}$, (c) $\Delta_{p_x} + e^{{\rm i} \Phi}\Delta_{p_y}$ and (d) $\Delta_{d_{x^2 -y^2}}+e^{{\rm i} \Phi}\Delta_{p_x}$.}
\label{Fig1} 
\end{figure}

In this section we provide details about the relative phase angle dependence of the total energy of various mixed order-parameter superconducting states. This analysis helps us in reducing the number of variational parameters used in our minimization scheme. Fig. \ref{Fig1} displays the results for the dependence of total energy on relative phase $\Phi$ between two order parameters. For $s$-wave and $d_{x^2-y^2}$ order with fixed magnitude of order parameters, we find that $\Phi = \pm \pi/2$ leads to the minimum energy for any value of chemical potential $\mu$ (see Fig. \ref{Fig1} ($a$)). Similarly, the relative phase angle between $p_x$ and $p_y$ order parameters, when both of them are assumed finite in magnitude, is $\pm \pi/2$ (see Fig. \ref{Fig1} ($c$)). On the other hand, the relative angle corresponding to the
minimum total energy $\Phi_{min}$ takes values $0$ or $\pi$ for $s$ and $d_{x^2-y^2}$, and $s$ and $p_x$ order parameters. These results do not depend on the choice of $\mu$ values.

\begin{figure}[H]
\includegraphics[width=\columnwidth]{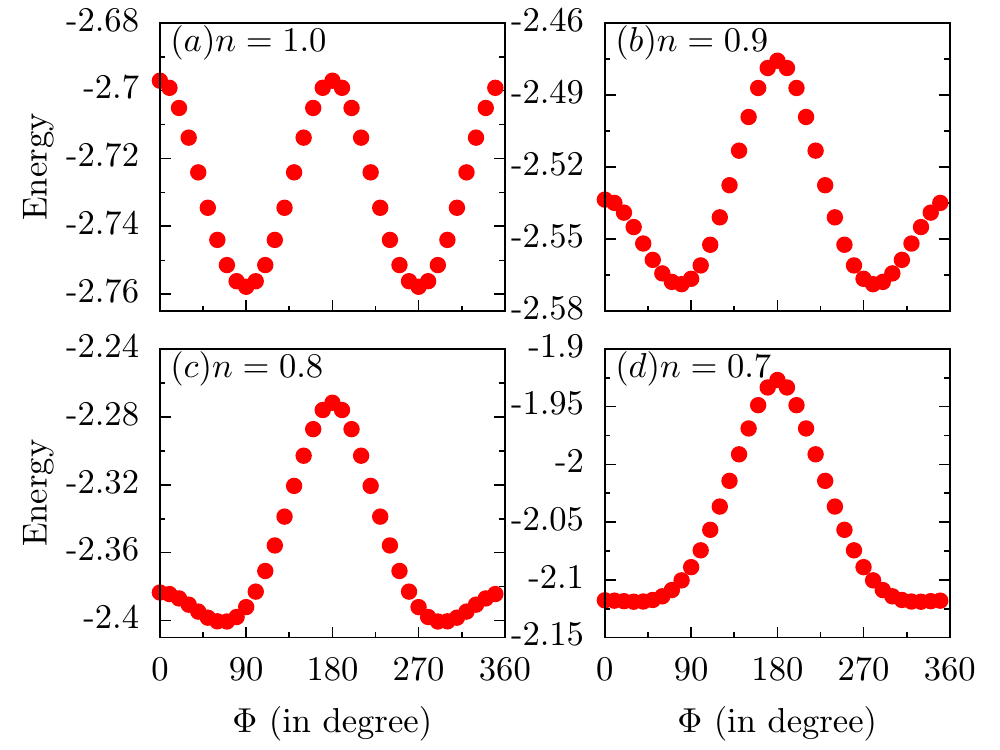}
\caption{ Variation of energy with relative phase angle ($\Phi$) between finite $\Delta_s$ and $\Delta_{d_{x^2 + y^2}}$ for different values of $\mu$: (a) $\mu=0$ ($n\approx1.0$), (b) $\mu=-0.2$ ($n\approx0.9$), (c) $\mu=-0.4$ ($n\approx0.8$) and (d) $\mu=-0.8$ ($n\approx0.7$).}
\label{Fig2}
\end{figure}

\begin{figure}[H]
\includegraphics[width= 0.98\columnwidth]{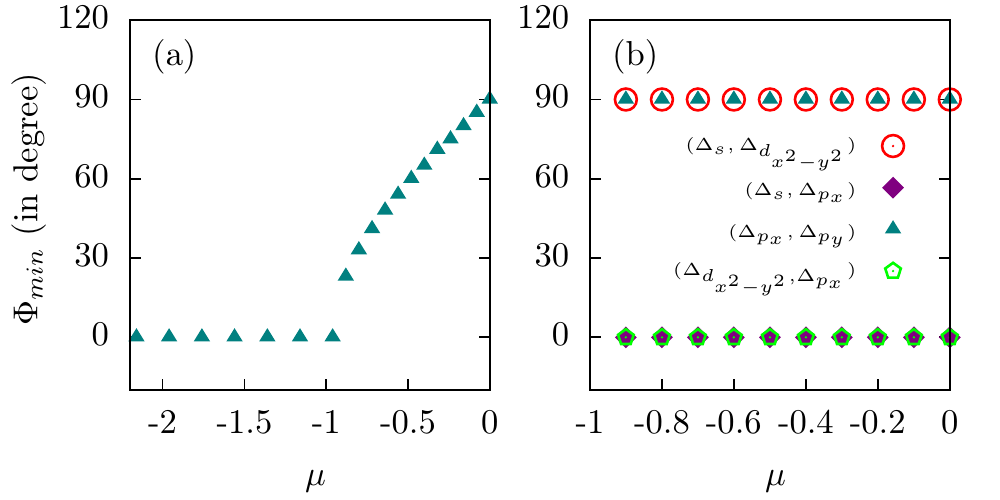}
\caption{ Variation of relative phase angle ($\Phi_{min}$) between two fnite orders, corresponding to minimum energy of the system, with chemical potential ($\mu$) for (a) $\Delta_s + e^{{\rm i}\Phi}\Delta_{d_{x^2+y^2}}$ and (b) rest of the possible pairings of such orders.}
\label{Fig3}
\end{figure}

\noindent
The relative angle between $s$-wave and $d_{x^2+y^2}$-wave order parameters shows an interesting behavior. $\Phi_{min}$ is found to evolve with change in $\mu$. For $\mu = 0$, corresponding to the 
half-filled band, $\Phi_{min} = \pi/2$. It decreases monotonically and becomes zero near $\mu = -1$, which corresponds to $n \approx 0.7$ (see Fig. \ref{Fig2}). These results are 
summarized in Fig. \ref{Fig3} wher we plot the Variation in $\Phi_{min}$ as a function of $\mu$. Clearly, the only order-parameter pair that shows a nontrivial variation of $\Phi_{min}$ is $s$-wave and 
$d_{x^2-y^2}$-wave order parameter pair.

\subsection{Self-consistency checks for the minimum energy solutions}
For completeness, we show here that various exotic superconducting states that are the minimum energy solutions of the mean-field Hamiltonian are stable solutions in terms of self-consistency.
Starting with initial values of $\{ \Delta \}$, we can recalculate different pair correlations using the following set of equations.

\begin{figure}
\includegraphics[width = 0.98\columnwidth]{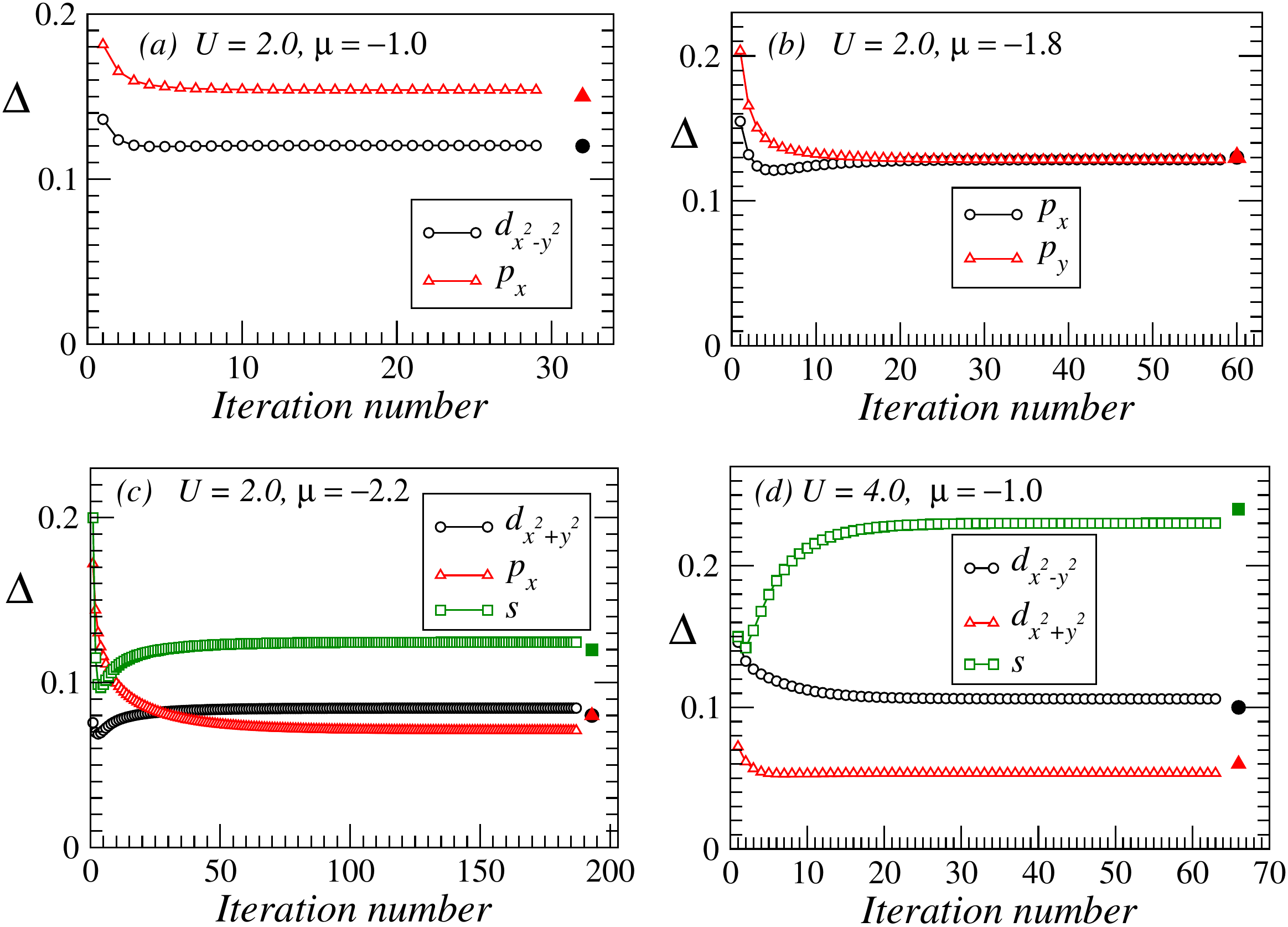}
\caption{Comparison of self-consistency results with minimization results: ($a$) variations of $\Delta_{d_{x^2-y^2}}$ and $\Delta_{p_x}$ with iteration number. ($b$) $\Delta_{p_x}$ and $\Delta_{p_y}$ as a function of iteration number. ($c$) $\Delta_{d_{x^2+y^2}}$, $\Delta_0$ and $\Delta_{p_x}$ as a function of iteration number, and ($d$) variations of $\Delta_{d_{x^2+y^2}}$, $\Delta_{x^2-y^2}$ and $\Delta_0$
with iterations. The filled symbols placed next to the $y$-axis denote the values of the corresponding parameters obtained within minimization approach presented in main text. The order parameters that are not shown here remain vanishingly small.}
\label{Fig4}
\end{figure}

\begin{eqnarray}
\Delta_0 & = & 1/N_s \sum_{\bf k} u_k v^*_k (2 f(E_{\bf k}) - 1)  \nonumber \\
\Delta^+_x & = & 1/N_s \sum_{\bf k} e^{{\rm i}k_x} ~u_k v^*_k (2 f(E_{\bf k}) - 1)  \nonumber \\
\Delta^-_x & = & 1/N_s \sum_{\bf k} e^{-{\rm i}k_x} ~u_k v^*_k (2 f(E_{\bf k}) - 1)  \nonumber \\
\Delta^+_y & = & 1/N_s \sum_{\bf k} e^{{\rm i}k_y} ~u_k v^*_k (2 f(E_{\bf k}) - 1)  \nonumber \\
\Delta^-_y & = & 1/N_s \sum_{\bf k} e^{-{\rm i}k_y} ~u_k v^*_k (2 f(E_{\bf k}) - 1).
\label{self-con}
\end{eqnarray}

\noindent
In the above, $E_{\bf k}$ are the quasiparticle eigenvalues, $f(E_{\bf k})$ denotes the Fermi function and $u_{\bf k}$, $v_{\bf k}$ are the coefficients that appear in the Bogoliubov transformation Eq. (\ref{BogT}) used in diagonalizing the mean-field Hamiltonian. We then repeat this procedure of diagonalizing the Hamiltonian for a given set $\{ \Delta \}$ and recalculating the set $\{ \Delta \}$ using Eq. (15) until the parameters converge within an accuracy limit of $10^{-6}$.

We show the results of the self-consistent calculations in Fig. \ref{Fig4}. The relevant order parameters are plotted as a function of the iteration number. In order to check the stability of the solutions 
we begin with the intial values of order parameters that are different from the values obtained via minimization. For the comparison we have picked representative parameter values corresponding to distinct phases shown in the phase diagram in main text. For example, Fig. \ref{Fig4}($a$) shows the results for $d_{x^2-y^2} + p_x$ state. We find that the values of these parameters converge very close to the values obtained in minimization approach. The filled symbols in each panel in Fig. \ref{Fig4} represent the corresponding values of parameters obtained via minimization. Similarly, for all other phases the values of the parameters obtained within self-consistent approach are match very well with those reported in minimization method.

\bibliographystyle {apsrev4-1}

\begin{thebibliography}{51}%
\makeatletter
\providecommand \@ifxundefined [1]{%
 \@ifx{#1\undefined}
}%
\providecommand \@ifnum [1]{%
 \ifnum #1\expandafter \@firstoftwo
 \else \expandafter \@secondoftwo
 \fi
}%
\providecommand \@ifx [1]{%
 \ifx #1\expandafter \@firstoftwo
 \else \expandafter \@secondoftwo
 \fi
}%
\providecommand \natexlab [1]{#1}%
\providecommand \enquote  [1]{``#1''}%
\providecommand \bibnamefont  [1]{#1}%
\providecommand \bibfnamefont [1]{#1}%
\providecommand \citenamefont [1]{#1}%
\providecommand \href@noop [0]{\@secondoftwo}%
\providecommand \href [0]{\begingroup \@sanitize@url \@href}%
\providecommand \@href[1]{\@@startlink{#1}\@@href}%
\providecommand \@@href[1]{\endgroup#1\@@endlink}%
\providecommand \@sanitize@url [0]{\catcode `\\12\catcode `\$12\catcode
  `\&12\catcode `\#12\catcode `\^12\catcode `\_12\catcode `\%12\relax}%
\providecommand \@@startlink[1]{}%
\providecommand \@@endlink[0]{}%
\providecommand \url  [0]{\begingroup\@sanitize@url \@url }%
\providecommand \@url [1]{\endgroup\@href {#1}{\urlprefix }}%
\providecommand \urlprefix  [0]{URL }%
\providecommand \Eprint [0]{\href }%
\providecommand \doibase [0]{http://dx.doi.org/}%
\providecommand \selectlanguage [0]{\@gobble}%
\providecommand \bibinfo  [0]{\@secondoftwo}%
\providecommand \bibfield  [0]{\@secondoftwo}%
\providecommand \translation [1]{[#1]}%
\providecommand \BibitemOpen [0]{}%
\providecommand \bibitemStop [0]{}%
\providecommand \bibitemNoStop [0]{.\EOS\space}%
\providecommand \EOS [0]{\spacefactor3000\relax}%
\providecommand \BibitemShut  [1]{\csname bibitem#1\endcsname}%
\let\auto@bib@innerbib\@empty
\bibitem [{\citenamefont {Tsuei}\ and\ \citenamefont
  {Kirtley}(2000)}]{Tsuei2000}%
  \BibitemOpen
  \bibfield  {author} {\bibinfo {author} {\bibfnamefont {C.~C.}\ \bibnamefont
  {Tsuei}}\ and\ \bibinfo {author} {\bibfnamefont {J.~R.}\ \bibnamefont
  {Kirtley}},\ }\href {\doibase 10.1103/RevModPhys.72.969} {\bibfield
  {journal} {\bibinfo  {journal} {Rev. Mod. Phys.}\ }\textbf {\bibinfo {volume}
  {72}},\ \bibinfo {pages} {969} (\bibinfo {year} {2000})}\BibitemShut
  {NoStop}%
\bibitem [{\citenamefont {Mathai}\ \emph {et~al.}(1995)\citenamefont {Mathai},
  \citenamefont {Gim}, \citenamefont {Black}, \citenamefont {Amar},\ and\
  \citenamefont {Wellstood}}]{Mathai1995}%
  \BibitemOpen
  \bibfield  {author} {\bibinfo {author} {\bibfnamefont {A.}~\bibnamefont
  {Mathai}}, \bibinfo {author} {\bibfnamefont {Y.}~\bibnamefont {Gim}},
  \bibinfo {author} {\bibfnamefont {R.~C.}\ \bibnamefont {Black}}, \bibinfo
  {author} {\bibfnamefont {A.}~\bibnamefont {Amar}}, \ and\ \bibinfo {author}
  {\bibfnamefont {F.~C.}\ \bibnamefont {Wellstood}},\ }\href {\doibase
  10.1103/PhysRevLett.74.4523} {\bibfield  {journal} {\bibinfo  {journal}
  {Phys. Rev. Lett.}\ }\textbf {\bibinfo {volume} {74}},\ \bibinfo {pages}
  {4523} (\bibinfo {year} {1995})}\BibitemShut {NoStop}%
\bibitem [{\citenamefont {Li}\ \emph {et~al.}(1993)\citenamefont {Li},
  \citenamefont {Koltenbah},\ and\ \citenamefont {Joynt}}]{Li1993}%
  \BibitemOpen
  \bibfield  {author} {\bibinfo {author} {\bibfnamefont {Q.~P.}\ \bibnamefont
  {Li}}, \bibinfo {author} {\bibfnamefont {B.~E.~C.}\ \bibnamefont
  {Koltenbah}}, \ and\ \bibinfo {author} {\bibfnamefont {R.}~\bibnamefont
  {Joynt}},\ }\href {\doibase 10.1103/PhysRevB.48.437} {\bibfield  {journal}
  {\bibinfo  {journal} {Phys. Rev. B}\ }\textbf {\bibinfo {volume} {48}},\
  \bibinfo {pages} {437} (\bibinfo {year} {1993})}\BibitemShut {NoStop}%
\bibitem [{\citenamefont {{Van Harlingen}}(1995)}]{VanHarlingen1995}%
  \BibitemOpen
  \bibfield  {author} {\bibinfo {author} {\bibfnamefont {D.~J.}\ \bibnamefont
  {{Van Harlingen}}},\ }\href {\doibase 10.1103/RevModPhys.67.515} {\bibfield
  {journal} {\bibinfo  {journal} {Rev. Mod. Phys.}\ }\textbf {\bibinfo {volume}
  {67}},\ \bibinfo {pages} {515} (\bibinfo {year} {1995})}\BibitemShut
  {NoStop}%
\bibitem [{\citenamefont {Muller}(1995)}]{Muller1995}%
  \BibitemOpen
  \bibfield  {author} {\bibinfo {author} {\bibfnamefont {K.~A.}\ \bibnamefont
  {Muller}},\ }\href {\doibase 10.1038/377133a0} {\bibfield  {journal}
  {\bibinfo  {journal} {Nature}\ }\textbf {\bibinfo {volume} {377}},\ \bibinfo
  {pages} {133} (\bibinfo {year} {1995})}\BibitemShut {NoStop}%
\bibitem [{\citenamefont {Betouras}\ and\ \citenamefont
  {Joynt}(1999)}]{Betouras1999}%
  \BibitemOpen
  \bibfield  {author} {\bibinfo {author} {\bibfnamefont {J.}~\bibnamefont
  {Betouras}}\ and\ \bibinfo {author} {\bibfnamefont {R.}~\bibnamefont
  {Joynt}},\ }\href {\doibase 10.1016/S0921-4534(99)00170-7} {\bibfield
  {journal} {\bibinfo  {journal} {Phys. C Supercond.}\ }\textbf {\bibinfo
  {volume} {317-318}},\ \bibinfo {pages} {669} (\bibinfo {year}
  {1999})}\BibitemShut {NoStop}%
\bibitem [{\citenamefont {Khasanov}\ \emph {et~al.}(2007)\citenamefont
  {Khasanov}, \citenamefont {Shengelaya}, \citenamefont {Maisuradze},
  \citenamefont {Mattina}, \citenamefont {Bussmann-Holder}, \citenamefont
  {Keller},\ and\ \citenamefont {M{\"{u}}ller}}]{Khasanov2007}%
  \BibitemOpen
  \bibfield  {author} {\bibinfo {author} {\bibfnamefont {R.}~\bibnamefont
  {Khasanov}}, \bibinfo {author} {\bibfnamefont {A.}~\bibnamefont
  {Shengelaya}}, \bibinfo {author} {\bibfnamefont {A.}~\bibnamefont
  {Maisuradze}}, \bibinfo {author} {\bibfnamefont {F.~L.}\ \bibnamefont
  {Mattina}}, \bibinfo {author} {\bibfnamefont {A.}~\bibnamefont
  {Bussmann-Holder}}, \bibinfo {author} {\bibfnamefont {H.}~\bibnamefont
  {Keller}}, \ and\ \bibinfo {author} {\bibfnamefont {K.~A.}\ \bibnamefont
  {M{\"{u}}ller}},\ }\href {\doibase 10.1103/PhysRevLett.98.057007} {\bibfield
  {journal} {\bibinfo  {journal} {Phys. Rev. Lett.}\ }\textbf {\bibinfo
  {volume} {98}},\ \bibinfo {pages} {057007} (\bibinfo {year}
  {2007})}\BibitemShut {NoStop}%
\bibitem [{\citenamefont {Armitage}\ \emph {et~al.}(2010)\citenamefont
  {Armitage}, \citenamefont {Fournier},\ and\ \citenamefont
  {Greene}}]{Armitage2010}%
  \BibitemOpen
  \bibfield  {author} {\bibinfo {author} {\bibfnamefont {N.~P.}\ \bibnamefont
  {Armitage}}, \bibinfo {author} {\bibfnamefont {P.}~\bibnamefont {Fournier}},
  \ and\ \bibinfo {author} {\bibfnamefont {R.~L.}\ \bibnamefont {Greene}},\
  }\href {\doibase 10.1103/RevModPhys.82.2421} {\bibfield  {journal} {\bibinfo
  {journal} {Rev. Mod. Phys.}\ }\textbf {\bibinfo {volume} {82}},\ \bibinfo
  {pages} {2421} (\bibinfo {year} {2010})}\BibitemShut {NoStop}%
\bibitem [{\citenamefont {Razzoli}\ \emph {et~al.}(2012)\citenamefont
  {Razzoli}, \citenamefont {Drachuck}, \citenamefont {Keren}, \citenamefont
  {Radovic}, \citenamefont {Plumb}, \citenamefont {Chang}, \citenamefont
  {Mesot},\ and\ \citenamefont {Shi}}]{Razzoli2012}%
  \BibitemOpen
  \bibfield  {author} {\bibinfo {author} {\bibfnamefont {E.}~\bibnamefont
  {Razzoli}}, \bibinfo {author} {\bibfnamefont {G.}~\bibnamefont {Drachuck}},
  \bibinfo {author} {\bibfnamefont {A.}~\bibnamefont {Keren}}, \bibinfo
  {author} {\bibfnamefont {M.}~\bibnamefont {Radovic}}, \bibinfo {author}
  {\bibfnamefont {N.~C.}\ \bibnamefont {Plumb}}, \bibinfo {author}
  {\bibfnamefont {J.}~\bibnamefont {Chang}}, \bibinfo {author} {\bibfnamefont
  {J.}~\bibnamefont {Mesot}}, \ and\ \bibinfo {author} {\bibfnamefont
  {M.}~\bibnamefont {Shi}},\ }\href {\doibase 10.1103/PhysRevLett.110.047004}
  {\  (\bibinfo {year} {2012}),\ 10.1103/PhysRevLett.110.047004},\ \Eprint
  {http://arxiv.org/abs/1207.3486} {arXiv:1207.3486} \BibitemShut {NoStop}%
\bibitem [{\citenamefont {Kallin}(2012)}]{Kallin2012}%
  \BibitemOpen
  \bibfield  {author} {\bibinfo {author} {\bibfnamefont {C.}~\bibnamefont
  {Kallin}},\ }\href {\doibase 10.1088/0034-4885/75/4/042501} {\bibfield
  {journal} {\bibinfo  {journal} {Reports Prog. Phys.}\ }\textbf {\bibinfo
  {volume} {75}},\ \bibinfo {pages} {042501} (\bibinfo {year}
  {2012})}\BibitemShut {NoStop}%
\bibitem [{\citenamefont {Kallin}\ and\ \citenamefont
  {Berlinsky}(2016)}]{Kallin2016}%
  \BibitemOpen
  \bibfield  {author} {\bibinfo {author} {\bibfnamefont {C.}~\bibnamefont
  {Kallin}}\ and\ \bibinfo {author} {\bibfnamefont {J.}~\bibnamefont
  {Berlinsky}},\ }\href {\doibase 10.1088/0034-4885/79/5/054502} {\bibfield
  {journal} {\bibinfo  {journal} {Reports Prog. Phys.}\ }\textbf {\bibinfo
  {volume} {79}},\ \bibinfo {pages} {054502} (\bibinfo {year}
  {2016})}\BibitemShut {NoStop}%
\bibitem [{\citenamefont {Scaffidi}\ \emph {et~al.}(2014)\citenamefont
  {Scaffidi}, \citenamefont {Romers},\ and\ \citenamefont
  {Simon}}]{Scaffidi2014}%
  \BibitemOpen
  \bibfield  {author} {\bibinfo {author} {\bibfnamefont {T.}~\bibnamefont
  {Scaffidi}}, \bibinfo {author} {\bibfnamefont {J.~C.}\ \bibnamefont
  {Romers}}, \ and\ \bibinfo {author} {\bibfnamefont {S.~H.}\ \bibnamefont
  {Simon}},\ }\href {\doibase 10.1103/PhysRevB.89.220510} {\bibfield  {journal}
  {\bibinfo  {journal} {Phys. Rev. B}\ }\textbf {\bibinfo {volume} {89}},\
  \bibinfo {pages} {220510} (\bibinfo {year} {2014})}\BibitemShut {NoStop}%
\bibitem [{\citenamefont {Luke}\ \emph {et~al.}(1998)\citenamefont {Luke},
  \citenamefont {Fudamoto}, \citenamefont {Kojima}, \citenamefont {Larkin},
  \citenamefont {Merrin}, \citenamefont {Nachumi}, \citenamefont {Uemura},
  \citenamefont {Maeno}, \citenamefont {Mao}, \citenamefont {Mori},
  \citenamefont {Nakamura},\ and\ \citenamefont {Sigrist}}]{Luke1998}%
  \BibitemOpen
  \bibfield  {author} {\bibinfo {author} {\bibfnamefont {G.~M.}\ \bibnamefont
  {Luke}}, \bibinfo {author} {\bibfnamefont {Y.}~\bibnamefont {Fudamoto}},
  \bibinfo {author} {\bibfnamefont {K.~M.}\ \bibnamefont {Kojima}}, \bibinfo
  {author} {\bibfnamefont {M.~I.}\ \bibnamefont {Larkin}}, \bibinfo {author}
  {\bibfnamefont {J.}~\bibnamefont {Merrin}}, \bibinfo {author} {\bibfnamefont
  {B.}~\bibnamefont {Nachumi}}, \bibinfo {author} {\bibfnamefont {Y.~J.}\
  \bibnamefont {Uemura}}, \bibinfo {author} {\bibfnamefont {Y.}~\bibnamefont
  {Maeno}}, \bibinfo {author} {\bibfnamefont {Z.~Q.}\ \bibnamefont {Mao}},
  \bibinfo {author} {\bibfnamefont {Y.}~\bibnamefont {Mori}}, \bibinfo {author}
  {\bibfnamefont {H.}~\bibnamefont {Nakamura}}, \ and\ \bibinfo {author}
  {\bibfnamefont {M.}~\bibnamefont {Sigrist}},\ }\href {\doibase 10.1038/29038}
  {\bibfield  {journal} {\bibinfo  {journal} {Nature}\ }\textbf {\bibinfo
  {volume} {394}},\ \bibinfo {pages} {558} (\bibinfo {year}
  {1998})}\BibitemShut {NoStop}%
\bibitem [{\citenamefont {Belin}\ and\ \citenamefont
  {Behnia}(1997)}]{Belin1997}%
  \BibitemOpen
  \bibfield  {author} {\bibinfo {author} {\bibfnamefont {S.}~\bibnamefont
  {Belin}}\ and\ \bibinfo {author} {\bibfnamefont {K.}~\bibnamefont {Behnia}},\
  }\href {\doibase 10.1103/PhysRevLett.79.2125} {\bibfield  {journal} {\bibinfo
   {journal} {Phys. Rev. Lett.}\ }\textbf {\bibinfo {volume} {79}},\ \bibinfo
  {pages} {2125} (\bibinfo {year} {1997})}\BibitemShut {NoStop}%
\bibitem [{\citenamefont {Lee}\ \emph {et~al.}(1997)\citenamefont {Lee},
  \citenamefont {Naughton}, \citenamefont {Danner},\ and\ \citenamefont
  {Chaikin}}]{Lee1997}%
  \BibitemOpen
  \bibfield  {author} {\bibinfo {author} {\bibfnamefont {I.~J.}\ \bibnamefont
  {Lee}}, \bibinfo {author} {\bibfnamefont {M.~J.}\ \bibnamefont {Naughton}},
  \bibinfo {author} {\bibfnamefont {G.~M.}\ \bibnamefont {Danner}}, \ and\
  \bibinfo {author} {\bibfnamefont {P.~M.}\ \bibnamefont {Chaikin}},\ }\href
  {\doibase 10.1103/PhysRevLett.78.3555} {\bibfield  {journal} {\bibinfo
  {journal} {Phys. Rev. Lett.}\ }\textbf {\bibinfo {volume} {78}},\ \bibinfo
  {pages} {3555} (\bibinfo {year} {1997})}\BibitemShut {NoStop}%
\bibitem [{\citenamefont {Lee}\ \emph {et~al.}(2001)\citenamefont {Lee},
  \citenamefont {Brown}, \citenamefont {Clark}, \citenamefont {Strouse},
  \citenamefont {Naughton}, \citenamefont {Kang},\ and\ \citenamefont
  {Chaikin}}]{Lee2001}%
  \BibitemOpen
  \bibfield  {author} {\bibinfo {author} {\bibfnamefont {I.~J.}\ \bibnamefont
  {Lee}}, \bibinfo {author} {\bibfnamefont {S.~E.}\ \bibnamefont {Brown}},
  \bibinfo {author} {\bibfnamefont {W.~G.}\ \bibnamefont {Clark}}, \bibinfo
  {author} {\bibfnamefont {M.~J.}\ \bibnamefont {Strouse}}, \bibinfo {author}
  {\bibfnamefont {M.~J.}\ \bibnamefont {Naughton}}, \bibinfo {author}
  {\bibfnamefont {W.}~\bibnamefont {Kang}}, \ and\ \bibinfo {author}
  {\bibfnamefont {P.~M.}\ \bibnamefont {Chaikin}},\ }\href {\doibase
  10.1103/PhysRevLett.88.017004} {\bibfield  {journal} {\bibinfo  {journal}
  {Phys. Rev. Lett.}\ }\textbf {\bibinfo {volume} {88}},\ \bibinfo {pages}
  {017004} (\bibinfo {year} {2001})}\BibitemShut {NoStop}%
\bibitem [{\citenamefont {Lee}\ \emph {et~al.}(2003)\citenamefont {Lee},
  \citenamefont {Chow}, \citenamefont {Clark}, \citenamefont {Strouse},
  \citenamefont {Naughton}, \citenamefont {Chaikin},\ and\ \citenamefont
  {Brown}}]{Lee2003}%
  \BibitemOpen
  \bibfield  {author} {\bibinfo {author} {\bibfnamefont {I.~J.}\ \bibnamefont
  {Lee}}, \bibinfo {author} {\bibfnamefont {D.~S.}\ \bibnamefont {Chow}},
  \bibinfo {author} {\bibfnamefont {W.~G.}\ \bibnamefont {Clark}}, \bibinfo
  {author} {\bibfnamefont {M.~J.}\ \bibnamefont {Strouse}}, \bibinfo {author}
  {\bibfnamefont {M.~J.}\ \bibnamefont {Naughton}}, \bibinfo {author}
  {\bibfnamefont {P.~M.}\ \bibnamefont {Chaikin}}, \ and\ \bibinfo {author}
  {\bibfnamefont {S.~E.}\ \bibnamefont {Brown}},\ }\href {\doibase
  10.1103/PhysRevB.68.092510} {\bibfield  {journal} {\bibinfo  {journal} {Phys.
  Rev. B}\ }\textbf {\bibinfo {volume} {68}},\ \bibinfo {pages} {092510}
  (\bibinfo {year} {2003})}\BibitemShut {NoStop}%
\bibitem [{\citenamefont {Oh}\ and\ \citenamefont {Naughton}(2004)}]{Oh2004}%
  \BibitemOpen
  \bibfield  {author} {\bibinfo {author} {\bibfnamefont {J.~I.}\ \bibnamefont
  {Oh}}\ and\ \bibinfo {author} {\bibfnamefont {M.~J.}\ \bibnamefont
  {Naughton}},\ }\href {\doibase 10.1103/PhysRevLett.92.067001} {\bibfield
  {journal} {\bibinfo  {journal} {Phys. Rev. Lett.}\ }\textbf {\bibinfo
  {volume} {92}},\ \bibinfo {pages} {067001} (\bibinfo {year}
  {2004})}\BibitemShut {NoStop}%
\bibitem [{\citenamefont {Gor'kov}\ and\ \citenamefont
  {Rashba}(2001)}]{Gorkov2001}%
  \BibitemOpen
  \bibfield  {author} {\bibinfo {author} {\bibfnamefont {L.~P.}\ \bibnamefont
  {Gor'kov}}\ and\ \bibinfo {author} {\bibfnamefont {E.~I.}\ \bibnamefont
  {Rashba}},\ }\href {\doibase 10.1103/PhysRevLett.87.037004} {\bibfield
  {journal} {\bibinfo  {journal} {Phys. Rev. Lett.}\ }\textbf {\bibinfo
  {volume} {87}},\ \bibinfo {pages} {037004} (\bibinfo {year}
  {2001})}\BibitemShut {NoStop}%
\bibitem [{\citenamefont {Sergienko}(2004)}]{Sergienko2004}%
  \BibitemOpen
  \bibfield  {author} {\bibinfo {author} {\bibfnamefont {I.~A.}\ \bibnamefont
  {Sergienko}},\ }\href {\doibase 10.1103/PhysRevB.69.174502} {\bibfield
  {journal} {\bibinfo  {journal} {Phys. Rev. B}\ }\textbf {\bibinfo {volume}
  {69}},\ \bibinfo {pages} {174502} (\bibinfo {year} {2004})}\BibitemShut
  {NoStop}%
\bibitem [{\citenamefont {Musaelian}\ \emph {et~al.}(1996)\citenamefont
  {Musaelian}, \citenamefont {Betouras}, \citenamefont {Chubukov},\ and\
  \citenamefont {Joynt}}]{Musaelian1996}%
  \BibitemOpen
  \bibfield  {author} {\bibinfo {author} {\bibfnamefont {K.~A.}\ \bibnamefont
  {Musaelian}}, \bibinfo {author} {\bibfnamefont {J.}~\bibnamefont {Betouras}},
  \bibinfo {author} {\bibfnamefont {A.~V.}\ \bibnamefont {Chubukov}}, \ and\
  \bibinfo {author} {\bibfnamefont {R.}~\bibnamefont {Joynt}},\ }\href
  {\doibase 10.1103/PhysRevB.53.3598} {\bibfield  {journal} {\bibinfo
  {journal} {Phys. Rev. B}\ }\textbf {\bibinfo {volume} {53}},\ \bibinfo
  {pages} {3598} (\bibinfo {year} {1996})}\BibitemShut {NoStop}%
\bibitem [{\citenamefont {Micnas}\ \emph {et~al.}(1988)\citenamefont {Micnas},
  \citenamefont {Ranninger},\ and\ \citenamefont {Robaszkiewicz}}]{Micnas1988}%
  \BibitemOpen
  \bibfield  {author} {\bibinfo {author} {\bibfnamefont {R.}~\bibnamefont
  {Micnas}}, \bibinfo {author} {\bibfnamefont {J.}~\bibnamefont {Ranninger}}, \
  and\ \bibinfo {author} {\bibfnamefont {S.}~\bibnamefont {Robaszkiewicz}},\
  }\href {\doibase 10.1088/0022-3719/21/6/009} {\bibfield  {journal} {\bibinfo
  {journal} {J. Phys. C Solid State Phys.}\ }\textbf {\bibinfo {volume} {21}},\
  \bibinfo {pages} {L145} (\bibinfo {year} {1988})}\BibitemShut {NoStop}%
\bibitem [{\citenamefont {Micnas}\ \emph {et~al.}(1990)\citenamefont {Micnas},
  \citenamefont {Ranninger},\ and\ \citenamefont {Robaszkiewicz}}]{Micnas1990}%
  \BibitemOpen
  \bibfield  {author} {\bibinfo {author} {\bibfnamefont {R.}~\bibnamefont
  {Micnas}}, \bibinfo {author} {\bibfnamefont {J.}~\bibnamefont {Ranninger}}, \
  and\ \bibinfo {author} {\bibfnamefont {S.}~\bibnamefont {Robaszkiewicz}},\
  }\href {\doibase 10.1103/RevModPhys.62.113} {\bibfield  {journal} {\bibinfo
  {journal} {Rev. Mod. Phys.}\ }\textbf {\bibinfo {volume} {62}},\ \bibinfo
  {pages} {113} (\bibinfo {year} {1990})}\BibitemShut {NoStop}%
\bibitem [{\citenamefont {Keller}\ \emph {et~al.}(2001)\citenamefont {Keller},
  \citenamefont {Metzner},\ and\ \citenamefont {Schollw?ck}}]{Keller2001a}%
  \BibitemOpen
  \bibfield  {author} {\bibinfo {author} {\bibfnamefont {M.}~\bibnamefont
  {Keller}}, \bibinfo {author} {\bibfnamefont {W.}~\bibnamefont {Metzner}}, \
  and\ \bibinfo {author} {\bibfnamefont {U.}~\bibnamefont {Schollw?ck}},\
  }\href {\doibase 10.1103/PhysRevLett.86.4612} {\bibfield  {journal} {\bibinfo
   {journal} {Phys. Rev. Lett.}\ }\textbf {\bibinfo {volume} {86}},\ \bibinfo
  {pages} {4612} (\bibinfo {year} {2001})}\BibitemShut {NoStop}%
\bibitem [{\citenamefont {Paiva}\ \emph {et~al.}(2004)\citenamefont {Paiva},
  \citenamefont {dos Santos}, \citenamefont {Scalettar},\ and\ \citenamefont
  {Denteneer}}]{Paiva2004}%
  \BibitemOpen
  \bibfield  {author} {\bibinfo {author} {\bibfnamefont {T.}~\bibnamefont
  {Paiva}}, \bibinfo {author} {\bibfnamefont {R.~R.}\ \bibnamefont {dos
  Santos}}, \bibinfo {author} {\bibfnamefont {R.~T.}\ \bibnamefont
  {Scalettar}}, \ and\ \bibinfo {author} {\bibfnamefont {P.~J.~H.}\
  \bibnamefont {Denteneer}},\ }\href {\doibase 10.1103/PhysRevB.69.184501}
  {\bibfield  {journal} {\bibinfo  {journal} {Phys. Rev. B}\ }\textbf {\bibinfo
  {volume} {69}},\ \bibinfo {pages} {184501} (\bibinfo {year}
  {2004})}\BibitemShut {NoStop}%
\bibitem [{\citenamefont {Anderson}(1975)}]{Anderson1975}%
  \BibitemOpen
  \bibfield  {author} {\bibinfo {author} {\bibfnamefont {P.~W.}\ \bibnamefont
  {Anderson}},\ }\href {\doibase 10.1103/PhysRevLett.34.953} {\bibfield
  {journal} {\bibinfo  {journal} {Phys. Rev. Lett.}\ }\textbf {\bibinfo
  {volume} {34}},\ \bibinfo {pages} {953} (\bibinfo {year} {1975})}\BibitemShut
  {NoStop}%
\bibitem [{\citenamefont {Montorsi}\ and\ \citenamefont
  {Campbell}(1996)}]{Montorsi1996}%
  \BibitemOpen
  \bibfield  {author} {\bibinfo {author} {\bibfnamefont {A.}~\bibnamefont
  {Montorsi}}\ and\ \bibinfo {author} {\bibfnamefont {D.~K.}\ \bibnamefont
  {Campbell}},\ }\href {\doibase 10.1103/PhysRevB.53.5153} {\bibfield
  {journal} {\bibinfo  {journal} {Phys. Rev. B}\ }\textbf {\bibinfo {volume}
  {53}},\ \bibinfo {pages} {5153} (\bibinfo {year} {1996})}\BibitemShut
  {NoStop}%
\bibitem [{\citenamefont {Capone}\ \emph {et~al.}(2002)\citenamefont {Capone},
  \citenamefont {Castellani},\ and\ \citenamefont {Grilli}}]{Capone2002}%
  \BibitemOpen
  \bibfield  {author} {\bibinfo {author} {\bibfnamefont {M.}~\bibnamefont
  {Capone}}, \bibinfo {author} {\bibfnamefont {C.}~\bibnamefont {Castellani}},
  \ and\ \bibinfo {author} {\bibfnamefont {M.}~\bibnamefont {Grilli}},\ }\href
  {\doibase 10.1103/PhysRevLett.88.126403} {\bibfield  {journal} {\bibinfo
  {journal} {Phys. Rev. Lett.}\ }\textbf {\bibinfo {volume} {88}},\ \bibinfo
  {pages} {126403} (\bibinfo {year} {2002})}\BibitemShut {NoStop}%
\bibitem [{\citenamefont {Singh}\ and\ \citenamefont
  {Scalettar}(1991)}]{Singh1991}%
  \BibitemOpen
  \bibfield  {author} {\bibinfo {author} {\bibfnamefont {R.~R.~P.}\
  \bibnamefont {Singh}}\ and\ \bibinfo {author} {\bibfnamefont {R.~T.}\
  \bibnamefont {Scalettar}},\ }\href {\doibase 10.1103/PhysRevLett.66.3203}
  {\bibfield  {journal} {\bibinfo  {journal} {Phys. Rev. Lett.}\ }\textbf
  {\bibinfo {volume} {66}},\ \bibinfo {pages} {3203} (\bibinfo {year}
  {1991})}\BibitemShut {NoStop}%
\bibitem [{\citenamefont {Gyorffy}\ \emph {et~al.}(1991)\citenamefont
  {Gyorffy}, \citenamefont {Staunton},\ and\ \citenamefont
  {Stocks}}]{Gyorffy1991a}%
  \BibitemOpen
  \bibfield  {author} {\bibinfo {author} {\bibfnamefont {B.~L.}\ \bibnamefont
  {Gyorffy}}, \bibinfo {author} {\bibfnamefont {J.~B.}\ \bibnamefont
  {Staunton}}, \ and\ \bibinfo {author} {\bibfnamefont {G.~M.}\ \bibnamefont
  {Stocks}},\ }\href {\doibase 10.1103/PhysRevB.44.5190} {\bibfield  {journal}
  {\bibinfo  {journal} {Phys. Rev. B}\ }\textbf {\bibinfo {volume} {44}},\
  \bibinfo {pages} {5190} (\bibinfo {year} {1991})}\BibitemShut {NoStop}%
\bibitem [{\citenamefont {Freericks}(1993)}]{Freericks1993a}%
  \BibitemOpen
  \bibfield  {author} {\bibinfo {author} {\bibfnamefont {J.~K.}\ \bibnamefont
  {Freericks}},\ }\href {\doibase 10.1103/PhysRevB.48.3881} {\bibfield
  {journal} {\bibinfo  {journal} {Phys. Rev. B}\ }\textbf {\bibinfo {volume}
  {48}},\ \bibinfo {pages} {3881} (\bibinfo {year} {1993})}\BibitemShut
  {NoStop}%
\bibitem [{\citenamefont {Allen}\ and\ \citenamefont
  {Tremblay}(2001)}]{Allen2001a}%
  \BibitemOpen
  \bibfield  {author} {\bibinfo {author} {\bibfnamefont {S.}~\bibnamefont
  {Allen}}\ and\ \bibinfo {author} {\bibfnamefont {A.-M.~S.}\ \bibnamefont
  {Tremblay}},\ }\href {\doibase 10.1103/PhysRevB.64.075115} {\bibfield
  {journal} {\bibinfo  {journal} {Phys. Rev. B}\ }\textbf {\bibinfo {volume}
  {64}},\ \bibinfo {pages} {075115} (\bibinfo {year} {2001})}\BibitemShut
  {NoStop}%
\bibitem [{\citenamefont {Arrachea}\ and\ \citenamefont
  {Aligia}(1999)}]{Arrachea1999}%
  \BibitemOpen
  \bibfield  {author} {\bibinfo {author} {\bibfnamefont {L.}~\bibnamefont
  {Arrachea}}\ and\ \bibinfo {author} {\bibfnamefont {A.~A.}\ \bibnamefont
  {Aligia}},\ }\href {\doibase 10.1103/PhysRevB.59.1333} {\bibfield  {journal}
  {\bibinfo  {journal} {Phys. Rev. B}\ }\textbf {\bibinfo {volume} {59}},\
  \bibinfo {pages} {1333} (\bibinfo {year} {1999})}\BibitemShut {NoStop}%
\bibitem [{\citenamefont {Mayr}\ \emph {et~al.}(2006)\citenamefont {Mayr},
  \citenamefont {Alvarez}, \citenamefont {Moreo},\ and\ \citenamefont
  {Dagotto}}]{Mayr2006}%
  \BibitemOpen
  \bibfield  {author} {\bibinfo {author} {\bibfnamefont {M.}~\bibnamefont
  {Mayr}}, \bibinfo {author} {\bibfnamefont {G.}~\bibnamefont {Alvarez}},
  \bibinfo {author} {\bibfnamefont {A.}~\bibnamefont {Moreo}}, \ and\ \bibinfo
  {author} {\bibfnamefont {E.}~\bibnamefont {Dagotto}},\ }\href {\doibase
  10.1103/PhysRevB.73.014509} {\bibfield  {journal} {\bibinfo  {journal} {Phys.
  Rev. B}\ }\textbf {\bibinfo {volume} {73}},\ \bibinfo {pages} {014509}
  (\bibinfo {year} {2006})}\BibitemShut {NoStop}%
\bibitem [{\citenamefont {Maiti}\ and\ \citenamefont
  {Hirschfeld}(2015)}]{Maiti2015}%
  \BibitemOpen
  \bibfield  {author} {\bibinfo {author} {\bibfnamefont {S.}~\bibnamefont
  {Maiti}}\ and\ \bibinfo {author} {\bibfnamefont {P.~J.}\ \bibnamefont
  {Hirschfeld}},\ }\href {\doibase 10.1103/PhysRevB.92.094506} {\bibfield
  {journal} {\bibinfo  {journal} {Phys. Rev. B}\ }\textbf {\bibinfo {volume}
  {92}},\ \bibinfo {pages} {094506} (\bibinfo {year} {2015})}\BibitemShut
  {NoStop}%
\bibitem [{\citenamefont {Hirsch}\ and\ \citenamefont
  {Scalapino}(1985)}]{Hirsch1985b}%
  \BibitemOpen
  \bibfield  {author} {\bibinfo {author} {\bibfnamefont {J.~E.}\ \bibnamefont
  {Hirsch}}\ and\ \bibinfo {author} {\bibfnamefont {D.~J.}\ \bibnamefont
  {Scalapino}},\ }\href {\doibase 10.1103/PhysRevB.32.117} {\bibfield
  {journal} {\bibinfo  {journal} {Phys. Rev. B}\ }\textbf {\bibinfo {volume}
  {32}},\ \bibinfo {pages} {117} (\bibinfo {year} {1985})}\BibitemShut
  {NoStop}%
\bibitem [{\citenamefont {Meintrup}\ \emph {et~al.}(1995)\citenamefont
  {Meintrup}, \citenamefont {Schneider},\ and\ \citenamefont
  {Beck}}]{Meintrup1995}%
  \BibitemOpen
  \bibfield  {author} {\bibinfo {author} {\bibfnamefont {T.}~\bibnamefont
  {Meintrup}}, \bibinfo {author} {\bibfnamefont {T.}~\bibnamefont {Schneider}},
  \ and\ \bibinfo {author} {\bibfnamefont {H.}~\bibnamefont {Beck}},\ }\href
  {\doibase 10.1209/0295-5075/31/4/008} {\bibfield  {journal} {\bibinfo
  {journal} {Europhys. Lett.}\ }\textbf {\bibinfo {volume} {31}},\ \bibinfo
  {pages} {231} (\bibinfo {year} {1995})}\BibitemShut {NoStop}%
\bibitem [{\citenamefont {Kuroki}\ \emph {et~al.}(1994)\citenamefont {Kuroki},
  \citenamefont {Kusakabe},\ and\ \citenamefont {Aoki}}]{Kuroki1994}%
  \BibitemOpen
  \bibfield  {author} {\bibinfo {author} {\bibfnamefont {K.}~\bibnamefont
  {Kuroki}}, \bibinfo {author} {\bibfnamefont {K.}~\bibnamefont {Kusakabe}}, \
  and\ \bibinfo {author} {\bibfnamefont {H.}~\bibnamefont {Aoki}},\ }\href
  {\doibase 10.1103/PhysRevB.50.575} {\bibfield  {journal} {\bibinfo  {journal}
  {Phys. Rev. B}\ }\textbf {\bibinfo {volume} {50}},\ \bibinfo {pages} {575}
  (\bibinfo {year} {1994})}\BibitemShut {NoStop}%
\bibitem [{\citenamefont {Scalapino}(2012)}]{Scalapino2012}%
  \BibitemOpen
  \bibfield  {author} {\bibinfo {author} {\bibfnamefont {D.~J.}\ \bibnamefont
  {Scalapino}},\ }\href {\doibase 10.1103/RevModPhys.84.1383} {\bibfield
  {journal} {\bibinfo  {journal} {Rev. Mod. Phys.}\ }\textbf {\bibinfo {volume}
  {84}},\ \bibinfo {pages} {1383} (\bibinfo {year} {2012})}\BibitemShut
  {NoStop}%
\bibitem [{\citenamefont {Huang}\ \emph {et~al.}(2013)\citenamefont {Huang},
  \citenamefont {Lai}, \citenamefont {Shi},\ and\ \citenamefont
  {Tsai}}]{Huang2013}%
  \BibitemOpen
  \bibfield  {author} {\bibinfo {author} {\bibfnamefont {W.-M.}\ \bibnamefont
  {Huang}}, \bibinfo {author} {\bibfnamefont {C.-Y.}\ \bibnamefont {Lai}},
  \bibinfo {author} {\bibfnamefont {C.}~\bibnamefont {Shi}}, \ and\ \bibinfo
  {author} {\bibfnamefont {S.-W.}\ \bibnamefont {Tsai}},\ }\href {\doibase
  10.1103/PhysRevB.88.054504} {\bibfield  {journal} {\bibinfo  {journal} {Phys.
  Rev. B}\ }\textbf {\bibinfo {volume} {88}},\ \bibinfo {pages} {054504}
  (\bibinfo {year} {2013})}\BibitemShut {NoStop}%
\bibitem [{\citenamefont {Mori}(1989)}]{Mori1989}%
  \BibitemOpen
  \bibfield  {author} {\bibinfo {author} {\bibfnamefont {H.}~\bibnamefont
  {Mori}},\ }\href {\doibase 10.1143/JPSJ.58.1394} {\bibfield  {journal}
  {\bibinfo  {journal} {J. Phys. Soc. Japan}\ }\textbf {\bibinfo {volume}
  {58}},\ \bibinfo {pages} {1394} (\bibinfo {year} {1989})}\BibitemShut
  {NoStop}%
\bibitem [{\citenamefont {DeGennes}(1999)}]{deGennes1999}%
  \BibitemOpen
  \bibfield  {author} {\bibinfo {author} {\bibfnamefont {P.-G.}\ \bibnamefont
  {DeGennes}},\ }\href@noop {} {\emph {\bibinfo {title} {{Superconductivity of
  metals and alloys}}}}\ (\bibinfo  {publisher} {Advanced Book Program, Perseus
  Books},\ \bibinfo {year} {1999})\BibitemShut {NoStop}%
\bibitem [{\citenamefont {Onari}\ \emph {et~al.}(2004)\citenamefont {Onari},
  \citenamefont {Arita}, \citenamefont {Kuroki},\ and\ \citenamefont
  {Aoki}}]{Onari2004}%
  \BibitemOpen
  \bibfield  {author} {\bibinfo {author} {\bibfnamefont {S.}~\bibnamefont
  {Onari}}, \bibinfo {author} {\bibfnamefont {R.}~\bibnamefont {Arita}},
  \bibinfo {author} {\bibfnamefont {K.}~\bibnamefont {Kuroki}}, \ and\ \bibinfo
  {author} {\bibfnamefont {H.}~\bibnamefont {Aoki}},\ }\href {\doibase
  10.1103/PhysRevB.70.094523} {\bibfield  {journal} {\bibinfo  {journal} {Phys.
  Rev. B}\ }\textbf {\bibinfo {volume} {70}},\ \bibinfo {pages} {094523}
  (\bibinfo {year} {2004})}\BibitemShut {NoStop}%
\bibitem [{\citenamefont {Onari}\ \emph {et~al.}(2005)\citenamefont {Onari},
  \citenamefont {Arita}, \citenamefont {Kuroki},\ and\ \citenamefont
  {Aoki}}]{Onari2005}%
  \BibitemOpen
  \bibfield  {author} {\bibinfo {author} {\bibfnamefont {S.}~\bibnamefont
  {Onari}}, \bibinfo {author} {\bibfnamefont {R.}~\bibnamefont {Arita}},
  \bibinfo {author} {\bibfnamefont {K.}~\bibnamefont {Kuroki}}, \ and\ \bibinfo
  {author} {\bibfnamefont {H.}~\bibnamefont {Aoki}},\ }\href {\doibase
  10.1143/JPSJ.74.2579} {\bibfield  {journal} {\bibinfo  {journal} {J. Phys.
  Soc. Japan}\ }\textbf {\bibinfo {volume} {74}},\ \bibinfo {pages} {2579}
  (\bibinfo {year} {2005})}\BibitemShut {NoStop}%
\bibitem [{\citenamefont {Huscroft}\ and\ \citenamefont
  {Scalettar}(1997)}]{Huscroft1997a}%
  \BibitemOpen
  \bibfield  {author} {\bibinfo {author} {\bibfnamefont {C.}~\bibnamefont
  {Huscroft}}\ and\ \bibinfo {author} {\bibfnamefont {R.~T.}\ \bibnamefont
  {Scalettar}},\ }\href {\doibase 10.1103/PhysRevB.55.1185} {\bibfield
  {journal} {\bibinfo  {journal} {Phys. Rev. B}\ }\textbf {\bibinfo {volume}
  {55}},\ \bibinfo {pages} {1185} (\bibinfo {year} {1997})}\BibitemShut
  {NoStop}%
\bibitem [{\citenamefont {Fischer}\ \emph {et~al.}(2007)\citenamefont
  {Fischer}, \citenamefont {Kugler}, \citenamefont {Maggio-Aprile},
  \citenamefont {Berthod},\ and\ \citenamefont {Renner}}]{Fischer2007}%
  \BibitemOpen
  \bibfield  {author} {\bibinfo {author} {\bibfnamefont {O.}~\bibnamefont
  {Fischer}}, \bibinfo {author} {\bibfnamefont {M.}~\bibnamefont {Kugler}},
  \bibinfo {author} {\bibfnamefont {I.}~\bibnamefont {Maggio-Aprile}}, \bibinfo
  {author} {\bibfnamefont {C.}~\bibnamefont {Berthod}}, \ and\ \bibinfo
  {author} {\bibfnamefont {C.}~\bibnamefont {Renner}},\ }\href {\doibase
  10.1103/RevModPhys.79.353} {\bibfield  {journal} {\bibinfo  {journal} {Rev.
  Mod. Phys.}\ }\textbf {\bibinfo {volume} {79}},\ \bibinfo {pages} {353}
  (\bibinfo {year} {2007})}\BibitemShut {NoStop}%
\bibitem [{\citenamefont {Balatsky}\ \emph {et~al.}(2006)\citenamefont
  {Balatsky}, \citenamefont {Vekhter},\ and\ \citenamefont
  {Zhu}}]{Balatsky2006}%
  \BibitemOpen
  \bibfield  {author} {\bibinfo {author} {\bibfnamefont {A.~V.}\ \bibnamefont
  {Balatsky}}, \bibinfo {author} {\bibfnamefont {I.}~\bibnamefont {Vekhter}}, \
  and\ \bibinfo {author} {\bibfnamefont {J.-X.}\ \bibnamefont {Zhu}},\ }\href
  {\doibase 10.1103/RevModPhys.78.373} {\bibfield  {journal} {\bibinfo
  {journal} {Rev. Mod. Phys.}\ }\textbf {\bibinfo {volume} {78}},\ \bibinfo
  {pages} {373} (\bibinfo {year} {2006})}\BibitemShut {NoStop}%
\bibitem [{\citenamefont {Taylor}\ and\ \citenamefont
  {Kallin}(2012)}]{Taylor2012}%
  \BibitemOpen
  \bibfield  {author} {\bibinfo {author} {\bibfnamefont {E.}~\bibnamefont
  {Taylor}}\ and\ \bibinfo {author} {\bibfnamefont {C.}~\bibnamefont
  {Kallin}},\ }\href {\doibase 10.1103/PhysRevLett.108.157001} {\bibfield
  {journal} {\bibinfo  {journal} {Phys. Rev. Lett.}\ }\textbf {\bibinfo
  {volume} {108}},\ \bibinfo {pages} {157001} (\bibinfo {year}
  {2012})}\BibitemShut {NoStop}%
\bibitem [{\citenamefont {Monthoux}\ and\ \citenamefont
  {Lonzarich}(1999)}]{Monthoux1999}%
  \BibitemOpen
  \bibfield  {author} {\bibinfo {author} {\bibfnamefont {P.}~\bibnamefont
  {Monthoux}}\ and\ \bibinfo {author} {\bibfnamefont {G.~G.}\ \bibnamefont
  {Lonzarich}},\ }\href {\doibase 10.1103/PhysRevB.59.14598} {\bibfield
  {journal} {\bibinfo  {journal} {Phys. Rev. B}\ }\textbf {\bibinfo {volume}
  {59}},\ \bibinfo {pages} {14598} (\bibinfo {year} {1999})}\BibitemShut
  {NoStop}%
\bibitem [{\citenamefont {Movshovich}\ \emph {et~al.}(1998)\citenamefont
  {Movshovich}, \citenamefont {Hubbard}, \citenamefont {Salamon}, \citenamefont
  {Balatsky}, \citenamefont {Yoshizaki}, \citenamefont {Sarrao},\ and\
  \citenamefont {Jaime}}]{Movshovich1998}%
  \BibitemOpen
  \bibfield  {author} {\bibinfo {author} {\bibfnamefont {R.}~\bibnamefont
  {Movshovich}}, \bibinfo {author} {\bibfnamefont {M.~A.}\ \bibnamefont
  {Hubbard}}, \bibinfo {author} {\bibfnamefont {M.~B.}\ \bibnamefont
  {Salamon}}, \bibinfo {author} {\bibfnamefont {A.~V.}\ \bibnamefont
  {Balatsky}}, \bibinfo {author} {\bibfnamefont {R.}~\bibnamefont {Yoshizaki}},
  \bibinfo {author} {\bibfnamefont {J.~L.}\ \bibnamefont {Sarrao}}, \ and\
  \bibinfo {author} {\bibfnamefont {M.}~\bibnamefont {Jaime}},\ }\href
  {\doibase 10.1103/PhysRevLett.80.1968} {\bibfield  {journal} {\bibinfo
  {journal} {Phys. Rev. Lett.}\ }\textbf {\bibinfo {volume} {80}},\ \bibinfo
  {pages} {1968} (\bibinfo {year} {1998})}\BibitemShut {NoStop}%
\bibitem [{\citenamefont {Brown}(2015)}]{Brown2015}%
  \BibitemOpen
  \bibfield  {author} {\bibinfo {author} {\bibfnamefont {S.}~\bibnamefont
  {Brown}},\ }\href {\doibase 10.1016/j.physc.2015.02.030} {\bibfield
  {journal} {\bibinfo  {journal} {Phys. C Supercond. its Appl.}\ }\textbf
  {\bibinfo {volume} {514}},\ \bibinfo {pages} {279} (\bibinfo {year}
  {2015})}\BibitemShut {NoStop}%
\end{thebibliography}

\end{document}